\documentclass[aip,reprint,nolongbibliography,superscriptaddress]{revtex4-2}
\usepackage[T1]{fontenc}
\usepackage[utf8]{inputenc}
\usepackage{graphicx}
\usepackage{amsmath}
\usepackage{xcolor}
\usepackage{txfonts}
\usepackage{times}
\usepackage{upgreek}
\usepackage{hyperref}
\usepackage{gensymb}
\usepackage{orcidlink}
\usepackage[normalem]{ulem}

\definecolor{greyed}{RGB}{170, 170, 170}
\definecolor{myblue}{RGB}{0, 40, 140}
\definecolor{blankcolor}{RGB}{220, 220, 230}
\hypersetup{
    unicode=true,          
    pdftitle={Active stabilization of kilogauss magnetic fields to the ppm level for magnetoassociation on ultranarrow Feshbach resonances},    
    pdfauthor={Mateusz Borkowski, et al.},     
    colorlinks=true,       
    linkcolor=myblue,          
    citecolor=myblue,        
    filecolor=myblue,      
    urlcolor=myblue           
}

\begin{document}
	\title{Active stabilization of kilogauss magnetic fields to the ppm level for magnetoassociation on ultranarrow Feshbach resonances}
 
	\author{Mateusz Borkowski\,\orcidlink{0000-0003-0236-8100}}
    \affiliation{Van der Waals-Zeeman Institute, Institute of Physics, University of Amsterdam, Science Park 904, 1098 XH Amsterdam, The Netherlands}
    \affiliation{Department of Physics, Columbia University, 538 West 120th Street, New York, NY 10027-5255, United States of America}
    \affiliation{Institute of Physics, Faculty of Physics, Astronomy and Informatics, Nicolaus Copernicus University, Grudziadzka 5, 87-100 Torun, Poland}
    
    \author{Lukas Reichsöllner\,\orcidlink{0000-0002-9986-2334}}
    \altaffiliation[Present address: ]{Planqc Ges.m.b.H, Lichtenbergstraße 8, 85748 Garching, Germany}
    \affiliation{Van der Waals-Zeeman Institute, Institute of Physics, University of Amsterdam, Science Park 904, 1098 XH Amsterdam, The Netherlands}

	\author{Premjith Thekkeppatt\,\orcidlink{0000-0002-1884-8398}} 	
    \affiliation{Van der Waals-Zeeman Institute, Institute of Physics, University of Amsterdam, Science Park 904, 1098 XH Amsterdam, The Netherlands}
    
    \author{Vincent Barbé\,\orcidlink{0000-0003-2414-9099}}
    \altaffiliation[Present address: ]{Department of Physics and Astronomy, LaserLaB, Vrije Universiteit Amsterdam, de Boelelaan 1081, 1081 HV Amsterdam, The Netherlands}	
    \affiliation{Van der Waals-Zeeman Institute, Institute of Physics, University of Amsterdam, Science Park 904, 1098 XH Amsterdam, The Netherlands}

    \author{Tijs van Roon}
    \affiliation{Technology Center, Faculty of Science, University of Amsterdam, Science Park 904, 1098 XH Amsterdam, The Netherlands}

	\author{Klaasjan van Druten\,\orcidlink{0000-0003-3326-9447}}	  
	\affiliation{Van der Waals-Zeeman Institute, Institute of Physics, University of Amsterdam, Science Park 904, 1098 XH Amsterdam, The Netherlands}
	
    \author{Florian Schreck\,\orcidlink{0000-0001-8225-8803}}
    \email[The author to whom correspondence may be addressed: ]{F.Schreck@uva.nl}
    \affiliation{Van der Waals-Zeeman Institute, Institute of Physics, University of Amsterdam, Science Park 904, 1098 XH Amsterdam, The Netherlands}
	    
	\date{\today}
	
\begin{abstract}
    Feshbach association of ultracold molecules using narrow resonances requires exquisite control of the applied magnetic field. Here we present a magnetic field control system to deliver magnetic fields of over 1000\,G with ppm-level precision integrated into an ultracold-atom experimental setup. We combine a battery-powered current-stabilized power supply with active feedback stabilization of the magnetic field using fluxgate magnetic field sensors. As a real-world test we perform microwave spectroscopy of ultracold Rb atoms and demonstrate an upper limit on our magnetic field stability of 2.4(3)\,mG at 1050~G [2.3(3)\,ppm relative] as determined from the spectral feature.
\end{abstract}

\maketitle

\begin{figure*}
    \includegraphics[width=\textwidth]{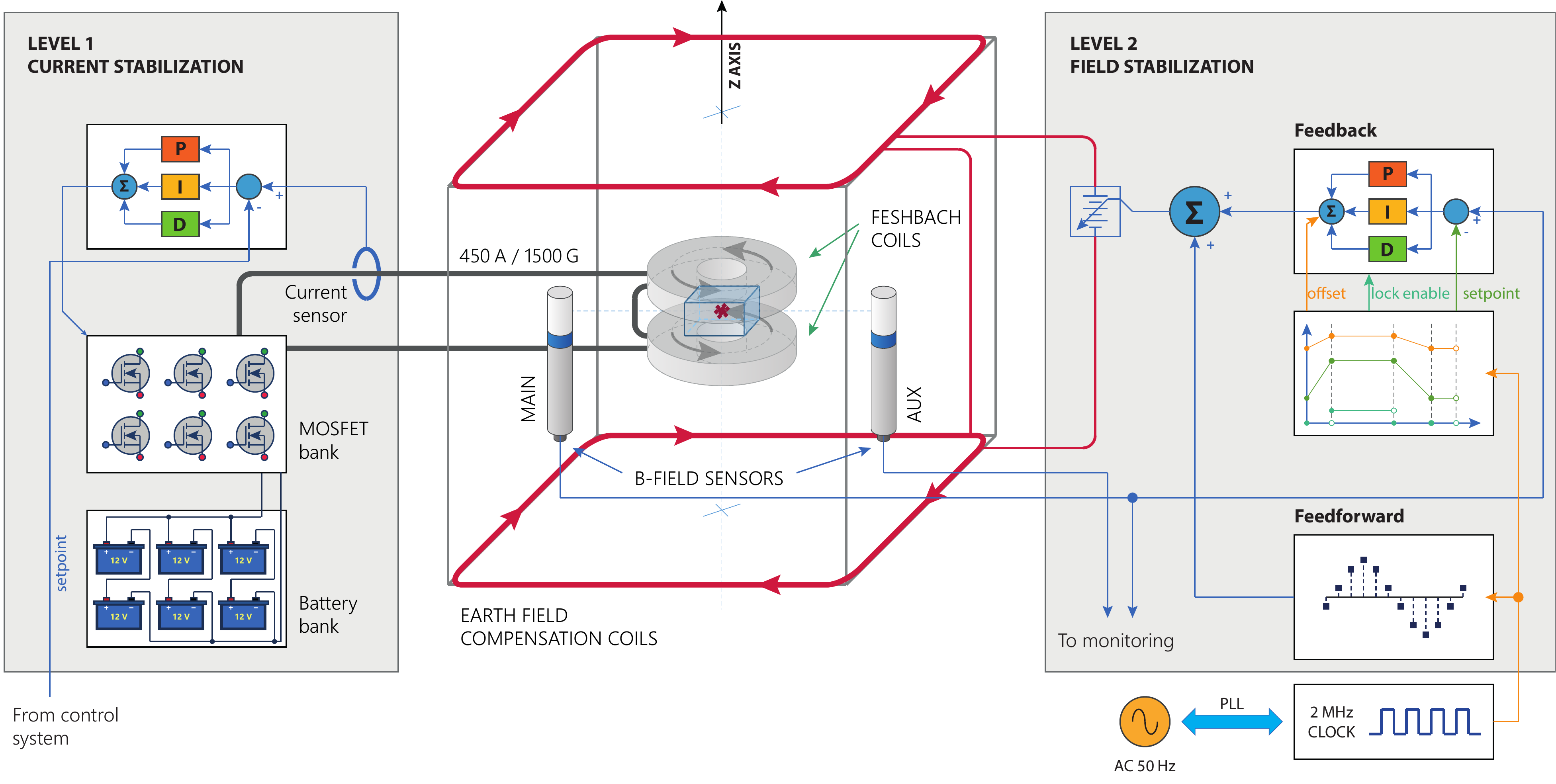}
    \caption{Schematic representation of the relevant magnetic field coils and fluxgate magnetic field sensors (not to scale). A glass science chamber is enclosed by two water-cooled Feshbach coils capable of withstanding over 450\,A of current and delivering up to $B_z \approx 1500\,$G of magnetic field at the atom position. The fluxgate field sensors are placed at a distance on two sides of the science chamber to avoid saturation from the large field generated by the Feshbach coils. Small field fluctuations are controlled by a set of large "Earth-Z" field compensation coils. The sensitivity of the two fluxgate sensors to changes in magnetic field from the Earth-Z coils is approximately 1:1, but less than 1\% of the Feshbach coil field reaches the sensors. The current reaching the Feshbach coils is actively stabilized with a PID loop fed by a current sensor ("Level 1", Sec.~\ref{sec:level1}). Additionally, we stabilize the environmental magnetic field in the laboratory via feedback and feedforward mechanisms ("Level 2", Sec.~\ref{sec:level2}). }
    \label{fig:coil_configuration}
\end{figure*}
 
\section{Introduction}
    Feshbach resonances\cite{Chin2010fri} have taken a prominent place in the ultracold atom toolbox. In a Feshbach resonance, the relative kinetic energy of a colliding atomic pair is tuned to the position of a molecular bound state of the two atoms. Typically these two states correspond to internal atomic states with different magnetic moments, and varying the external magnetic field is the most common way to achieve the required energy tuning. Magnetically tuned Feshbach resonances\cite{Inouye1998} have become a powerful tool in ultracold atom physics and quantum gas research\cite{Bloch2008mbp}, enabling tuning of the interactions\cite{Cornish2000,Roberts2001,Donley2001}, exploring the BEC-BCS crossover\cite{Parish2014bcs}, magnetoassociation of atoms to molecules\cite{Jochim2003,Kohler2006poc}, and controlling atom-ion collisions\cite{Weckesser2021ofr}. Furthermore, high-precision Feshbach resonance spectroscopy is promising for further applications in fundamental physics such as exploring Efimov resonances\cite{Greene2017ufp,Ferlaino2011eru}, and searching for the variation of fundamental constants\cite{Chin2006esf}.
    
    A particular challenge to magnetic field control arises when the coupling between the two states is weak and the consequently narrow Feshbach resonances are located at a high field, so that very accurate and precise control at this field strength is needed. For instance, our group has previously identified and observed several Feshbach resonances in ultracold thermal mixtures of Rb and Sr atoms at fields of several hundred gauss~\cite{Barbe2018ofr, Ciamei2018trs}. In principle, these could be used to coherently produce samples of ultracold ground-state RbSr molecules in a manner similar to alkali systems~\cite{Regal2003, Herbig2003, Ni2008, Danzl2010, Takekoshi2014a, Moses2015, Reichsollner2017, Voges2020} --- by Feshbach magnetoassociation of free atoms to weakly bound molecular states followed by a stimulated rapid adiabatic passage\cite{Vitanov2017} to the rovibrational ground state. Our interest in producing RbSr molecules is fueled by their tantalizing property of a $^2\Sigma$ ground state, with both a magnetic and electric dipole moment (in the molecular frame). The two inherent dipole moments could find use as independent ``tuning knobs'' increasing flexibility in such applications as many-body physics\cite{Bloch2008mbp}, as well as quantum computation\cite{Demille2002}, and simulation\cite{Micheli2006, Lewenstein2006}. 
    
    Realizing magnetoassociation of doublet molecules like RbSr has proven an elusive goal so far, although substantial progress has been made \cite{Barbe2018ofr, Ciamei2018trs, Guttridge2018, Green2019, Green2020, Franzen2022}. Because of the singlet character of Sr and other divalent atoms, the usual spin-spin coupling expected for a bialkali system is absent and the Feshbach resonances are extremely narrow~\cite{Zuchowski2010urs, Brue2012, Brue2013}. In fact, the most promising resonances for the RbSr system are located at 1313~G (for the Bose-Bose $^{87}$Rb+$^{84}$Sr system) and $519$~G (for the $^{87}$Rb +$^{87}$Sr Bose-Fermi mixture), and have respective widths of 1.7~mG and 16~mG \cite{Ciamei2018trs}. Thus, ppm-level control of the applied magnetic field is needed. Furthermore, the initial laser-cooling stages require switching between near-zero and quadrupolar magnetic fields and hence permanent magnets and other magnetic materials need to be avoided. Taken together, the levels of control and reproducibility desired for the magnetic fields pose a serious experimental challenge.
    
    Previously reported schemes either stabilized the Feshbach coil current or the environmental magnetic field in the laboratory, but not both. For example, low-noise current drivers for atomic physics averaging to sub-ppm precision were previously demonstrated~\cite{Yang2019uln, Thomas2020dig}. However, merely stabilizing the current does not guarantee an equally stable magnetic field at the atom position due to the influence of other sources of varying environmental magnetic fields. Therefore a number of active magnetic field stabilization schemes were also demonstrated. For example, Xu~\emph{et al.} demonstrated ppm-level magnetic field control at a 15-G level \cite{Xu2019ulnmag}, while Merkel~\emph{et al.} showed sub-ppm level control at 150\,G in the context of ion trapping\cite{Merkel2019magstab}. Duan~\emph{et al.} were able to extend coherence in ultracold Rb at low fields~\cite{Duan2022}. It is important to note that active magnetic field control typically involves both the compensation of slow field drifts (via a feedback mechanism), as well as the cancellation of oscillatory fields created by mains power (usually achieved via a feedforward). A number of adaptive feedforward schemes has been reported~\cite{ODwyer2020, Pyragius2021, Wei2022}

    Here we present a dual-level magnetic field control system designed to meet these challenges. It is generally not feasible to directly stabilize large magnetic fields on the order of 1000~G for lack of precise magnetic field sensors at this range and with the desired precision, so we employ a hybrid approach, where we actively stabilize both the current through the magnetic-field (``Feshbach'') coils, as well as the environmental magnetic field as measured by external sensors. We demonstrate the desired ppm-level control of $\sim$\,kG magnetic fields by performing microwave spectroscopy of ultracold Rb atoms. In the following sections we first describe the design and implementation of the two levels of control system. This is followed by the characterization of the performance, both using fluxgate magnetic field sensors and cold-atom magnetometry. Our system achieves a resolution of 2.4~mG at 1050~G (2.3 ppm relative), which to the best of our knowledge constitutes the most accurate kilogauss magnetic field control system to date.

\section{Active magnetic field control}

    A schematic overview of the system is shown in Fig.~\ref{fig:coil_configuration}. Our system comprises two levels of magnetic field control. At the first level, we stabilize the current through the Feshbach coils. The second level is designed to compensate the remaining residual magnetic field variation at the location of the atoms based on the measurement of the ambient magnetic field around the setup. We choose to only stabilize along the high-field axis; the other, orthogonal, magnetic field components add in quadrature to the high fields along this axis, and have negligible effect in practice.

    We also note one additional source of instability: the temperature of the Feshbach coil. Even though the coil is water-cooled, passing large currents of several hundred amps causes the coil to heat up over time by about 1\degree{}C on average. The associated thermal distortion causes up to 20~mG shift on the magnetic field value. To mitigate this we simply run our experiment continuously for about half an hour before taking data that requires high magnetic field stability.

\subsection{Level 1: Control of coil current\label{sec:level1}}
    The first level of control is based on stabilizing the current supplied to the Feshbach coils. In short, the elements are as follows: a battery stack is used as the low-noise power source. The current supplied to the coils is varied using a MOSFET bank, and is measured using a current transducer. The system is controlled by combining a proportional, integral and derivative (PID) controller and feed-forward signals. The time-varying setpoint is provided by the overall system that drives the full quantum gas experiment \cite{Schreck2011control}. We now describe each of these elements in more detail, and then give some specifics on the operation procedure and achieved stability.

    The Feshbach coil design follows that described in a previous work\cite{Wille2009poo}. The coils are water cooled and can withstand over 450\,A of current, delivering up to $B_z\approx 1500$\,G of magnetic field at the atom position between the two coils. The coils have an estimated inductance of 150\,$\mu$H and a combined DC resistance of 27\,m$\Omega$. The battery stack consists of six absorbent-glass-mat (AGM) deep cycle batteries as a power source (Victron BAT412151084), each rated at 12V nominal voltage and 165 Ah capacity. To achieve a 24V nominal voltage we first connect two batteries each in series. Then we connect the resulting three pairs in parallel to obtain the capability to deliver a maximum current of 450\,A and a capacity of 495\,Ah. By default the battery stack is charged in constant-voltage-charging mode at 27.3\,V using two chargers in series for improved balance. While running the experimental sequence we use a pair of high current relays (Schaltbau C600) to alternate between charging the batteries and powering the Feshbach coils. While charging, the high current system is galvanically isolated from the experimental setup to avoid noise. The water-cooled MOSFET bank consists of 5 banks of 20 balanced parallelized MOSFETs (IXYS IXTK 140N20P) each with a $0.05\,\Omega$ shunt resistor\cite{horowitz2015art} at its source terminal. We employ a standard MOSFET linearization technique: the shunt resistor acts as a sensor for the current passing through the MOSFET and Feshbach coils and provides feedback to a high-precision operational amplifier (OPA277P). The amplifier output automatically controls the MOSFET gate terminal such that the voltage on the shunt resistor matches the control voltage provided by a DAC. Voltage oscillations at the gate, sometimes referred to as gate ringing, are reduced by an additional resistor. We suppress the Miller effect~\cite{Miller1920dot} by reducing the bandwidth to about 300 Hz using an analog filter. For every group of 10 MOSFET drivers we use one differential amplifier to produce an error signal for the feedback loop. A snubber circuit across both coils was also added to suppress oscillations even further.
    
    We actively stabilize the coil current by monitoring its value using a current transducer (LEM ITN 600-S ULTRASTAB), which we expect to allow current stabilization at the 10 ppm level or better. The current flow is actively stabilized using a field-programmable gate array (Xilinx Zynq-7000) implementing the PID and a hard-coded feed-forward (not shown in Fig.~\ref{fig:coil_configuration}) signal processing to drive the linearized MOSFET bank. The current signal is digitized using a 24-bit ADC (Analog Devices AD7763); a 20-bit DAC (Analog Devices AD5791) is used to produce the analog signal to drive the MOSFETs. For both the ADC and DAC we use commercially available evaluation boards with the following modifications: the input stage of the ADC board was replaced to match the bandwidth of the current transducer; the on-board voltage reference was replaced with an external one and finally the shunt resistor for the current transducer was placed directly on-board. Both the ADC and DAC boards were placed in separate shielded containers and each had its own highly-stable low-dropout-regulator power supply. The ADC is running at a relatively high sample rate (78 kHz) and this is being downsampled via an FIR filter by the FPGA.

    The overall system has a relatively large bandwidth, $\approx 0-100$\,Hz, and 1.7-ppm relative stability, as measured over 10 minutes, at 80\% of the maximum output, i.e. around 1300~G magnetic field. The power-supply design is more generally applicable, not only to experiments involving similarly narrow Feshbach resonances, but to any setup with similarly extreme requirements for stability and control of an electric current. 
 
    For longer measurements of several hours this system alone achieves a 5-10 mG stability (see top panel of Fig. \ref{fig:sweepspectra} below). To achieve greater stability, we need to account and compensate for fluctuations in other, environmental, magnetic field sources. This is achieved by the second level of control, described next.

\begin{figure*}
    \includegraphics[width=0.95\textwidth]{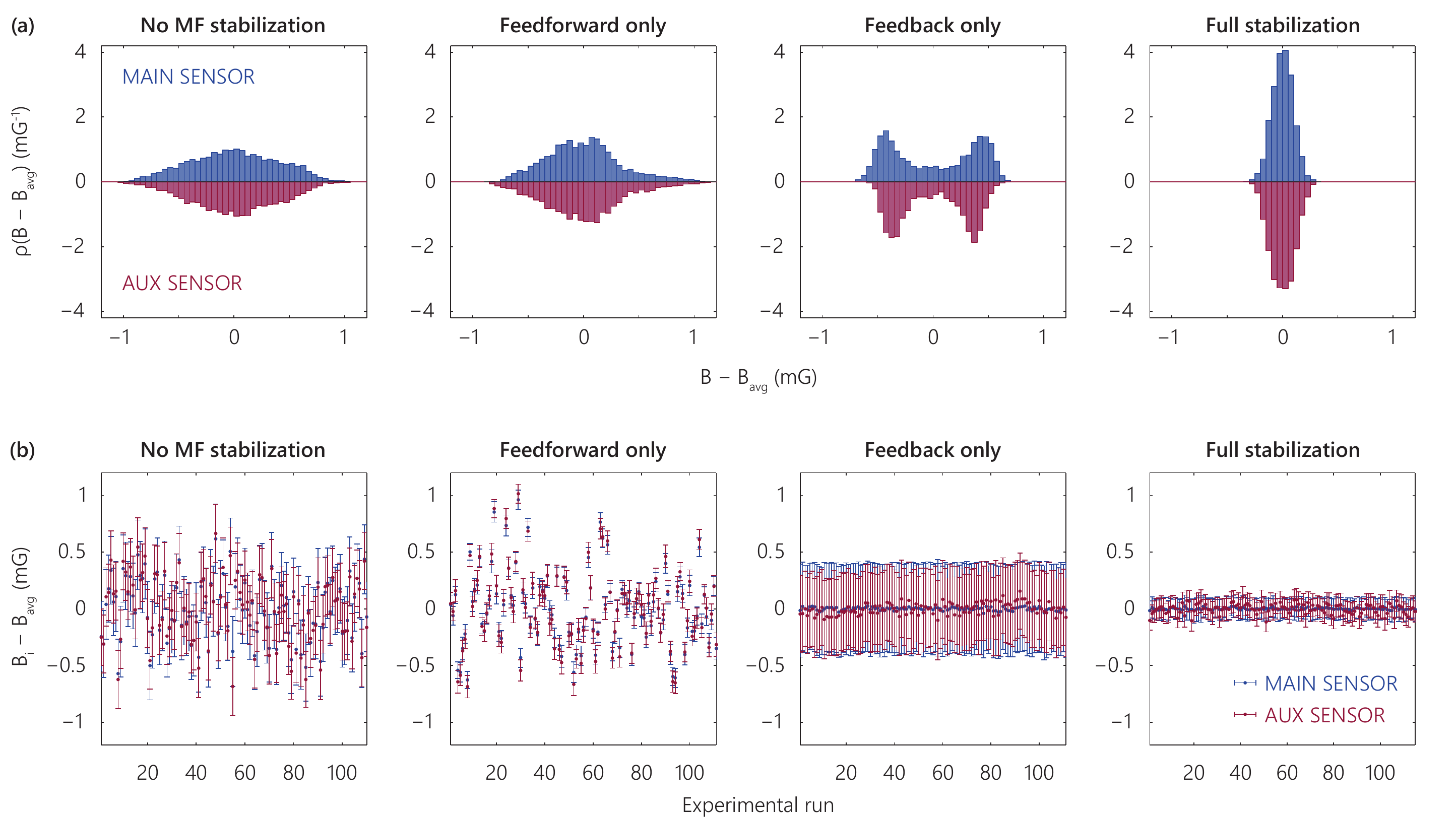}
    \caption{
        Testing the stability of the magnetic field using fluxgate sensors. 
        Panel (a) shows histograms of the magnetic fields experienced by the MAIN and AUX fluxgate sensors (Fig.~\ref{fig:coil_configuration}). Of particular interest is the bimodal distribution seen for the ``Feedback only'' case consistent with a mains sinusoidal signal; adding an out-of-phase feedforward signal effectively removes this interference.
        Panel (b) depicts magnetic fields over individual experimental runs of 100-ms duration. Each data point is the mean magnetic field $B_i$ for a given run with respect to the average field for all runs $B_{\rm avg}$; the error bar denotes the standard deviation of the field over the 100-ms run. Thus, the scatter in $B_i$ values is a measure of shot-to-shot magnetic field drift and the error bar provides information on the AC mains signal.
    }
    \label{fig:histograms}
\end{figure*}
     
\subsection{Level 2: Control of environmental magnetic fields \label{sec:level2}}
    Stable Feshbach coil current does not guarantee an equally stable magnetic field at the atom position as there are other varying magnetic fields in the laboratory. For us the two main sources of instability of the enviromental magnetic field are slow, seemingly random drifts of the magnetization in the laboratory and the oscillatory magnetic fields (nominally 50 Hz in Europe) created by the mains cables as current passes through them. The former are typically on the order of 1--2 mG during the day, likely due to thermal fluctuations of the magnetization in the room, but can be increased to a few mG due to the operation of other experiments in the building. On the other hand, the oscillatory magnetic fields are smaller -- well below 1~mG in amplitude -- but could potentially thwart our future efforts to magnetoassociate RbSr molecules by violating the adiabaticity condition~\cite{Mies2000, Goral2004,Chin2010fri}.
    
    To monitor these extra environmental fields we use a pair of fluxgate magnetic field sensors (Bartington Mag-03). The sensors are placed on two sides of our science chamber (MAIN and AUX in Fig.~\ref{fig:coil_configuration}). Since the sensors are capped at 10~G, we are forced to place them at a distance of 50~cm to avoid saturation or even destruction with the large Feshbach fields. The Feshbach field decreases roughly as distance cubed and at the sensor position is reduced by over two orders of magnitude.  

    We combat the environmental magnetic field fluctuations with a set of Earth field compensation coils in a near-Helmholtz configuration. The coils are rectangular in shape, with sides 140~cm and 100~cm long, are placed 90~cm apart and easily encompass both the science chamber and the two magnetic field sensors to ensure good uniformity of the magnetic field they produce. The coils are driven by a commercial laboratory power supply (Elektro-Automatik EA-PS 3016-10) controlled by a custom signal processing unit. Our power supply can pass up to 10~A of current and we typically run at close to 5~A for maximum flexibility. At the atom position, increasing the coil current by 1~A induces an additional field of 159~mG in the same direction as the Feshbach field, as determined from RF spectroscopy of ground-state Rb atoms. The same change in current causes the MAIN sensor reading to increase by 140~mG. Thus, the two match to better than 15\%.
    
    Our signal processing scheme is based around a 32-bit ARM SAM3X microcontroller (Arduino Due). The output voltages of the two magnetic field sensors are read out using a 24-bit analog to digital converter (Analog Devices AD7175-8) operating at a 1~kHz sample rate to achieve a peak-to-peak resolution above 20 bits ($<10^{-6}$). The readout is interleaved between the two sensors giving an effective sample rate of 500~Hz per sensor and an upper limit on the servo bandwidth of 250~Hz. The measured voltages are additionally sent over Arduino Due's native USB port to one of the experiment control computers for monitoring. The control voltage for the Earth field coils is produced by a 20-bit digital-to-analog converter (Analog Devices AD5791). Unlike Level~1 electronics here we have used custom designed boards for both the ADC and DAC portions of the system. The microcontroller communicates with the ADC and DAC chips over the SPI bus. The timing of the experimental sequence relies on a 2\,MHz oscillator synchronized via a phase-locked-loop (PLL) to the 50\,Hz mains, which we found can fluctuate up to $\pm 1$\%. The ADC, however, relies on having a stable 1 kHz sample rate. To align the two, we feed the 2 MHz signal into a 16-bit counter (a pair of 74HC590). On each new sample from the ADC, the counter is read out by the microcontroller to find out the current position in the experimental sequence.

    Prior to the development of the present magnetic field control scheme, the Earth field compensation coil has been a vital part of the sample preparation sequence to e.g. control the position the atom cloud of a magneto-optical trap with respect to our optical dipole trap. To maintain this functionality, we now, prior to launching the experimental sequence, send a list of control points to the microcontroller over the native USB bus (right hand side of Fig.~\ref{fig:coil_configuration}). Each point defines an \emph{offset} -- a baseline current to be sent into the coil -- and a \emph{setpoint}: the target magnetic field for the PID loop ("Feedback" in Fig.~\ref{fig:coil_configuration}) to stabilize to. Finally, each point also has a \emph{lock} flag that determines whether or not the PID loop is engaged. The microcontroller then linearly interpolates the \emph{offset} and \emph{setpoint} control points for a given time in the sequence. This allows us to use just the \emph{offset} signals to reproduce the prior experimental sequence (without engaging the \emph{lock}). Once we reach a point where the magnetic field stability is necessary, e.g. for Feshbach association, we can engage the \emph{lock} and activate the PID loop. We typically still use the \emph{offset} signal to bring the baseline magnetic field as close as possible to the desired value and we only use the PID loop for the remaining few tens of mG. Finally, it is worth noting that since both the \emph{setpoint} and \emph{offset} are interpolated, we are able to use the PID stabilization loop on top of linear field ramps necessary for e.g. magnetoassociation of Feshbach molecules.

    To compensate for oscillatory magnetic fields we synthesize an 180$^{\degree}$ out-of-phase 50~Hz sinewave. This is done within our signal processing microcontroller using a software sinewave generator. Its phase is aligned with the position in the sequence as determined by the PLL-locked 2-MHz clock signal and therefore is automatically in sync with the AC mains line. We set the amplitude and relative phase of the synthesized signal by hand to minimize the amplitude of the remaining interference. We find that we can reduce the RMS amplitude of oscillatory magnetic fields from about 270--300~$\mu$G down to 60--100~$\mu$G depending on the day. This depends, however, on the placement of any transformer based power supplies close to the experimental setup and a large part of minimizing this interference was to move many of the power supplies out of the laboratory and to rely on low voltage DC rails for power. While in principle we could also compensate for higher harmonics, e.g. 150 Hz, we found these to have worse spatial uniformity and day-to-day stability than the fundamental so we dismissed that as impractical.

    The signal flow proceeds as follows: the analog signals from the two fluxgate sensors are digitized by the ADC. Both signals are sent over Arduino Due's native USB port to a lab computer for monitoring. The signal from the MAIN sensor is compared with the sequencer's \emph{setpoint} signal to produce an error signal for the Feedback PID. The servo signal produced by the PID is summed together with the \emph{offset} signal from the sequencer. The result is then summed with the output of the sinewave generator to produce the final control signal. This is converted back into analog voltage using the DAC and sent to the power supply for the Earth field compensation coil.
    
\section{Testing the magnetic field stabilization}

    We test our magnetic field stabilization scheme in two different ways. First, we use the auxiliary magnetic field sensors to test our ability to stabilize the environmental magnetic field. Second, we use microwave spectroscopy of ultracold Rb atoms to perform magnetometry of the total magnetic field generated at the position of the atomic sample in the science chamber.

\subsection{Fluxgate sensors}

    We first test our ability to stabilize the environmental interference by use of the fluxgate sensors on the two sides of the experimental chamber. To do so, we run a realistic experimental sequence where we ramp our Feshbach field coils to 1050\,G, let the magnetic fields stabilize and then engage the selected stabilization mechanisms: feedforward and/or feedback. Then, we record the magnetic field signals from both sensors over a 100\,ms time window: so chosen because it matches our expected experimental time scales (trapped molecules are expected to have short lifetimes) and also because it covers five full 50\,Hz AC cycles. We repeated this experiment over a hundred times each for all combinations of the feedforward and feedback being off or on.

    Figure~\ref{fig:histograms}~(a) shows histograms of the recorded sensor signals lumped together from all experimental runs. If no environmental stabilization is engaged, both sensors show a roughly Gaussian distribution of magnetic field values, with a standard deviation $\sigma$ of 0.40\,mG and 0.38\,mG for the MAIN and AUX sensors, respectively. The two datasets are strongly correlated, with Pearson's correlation coefficient of 94\% showing that the environmental contribution to the magnetic field is to a large extent spatially uniform. This gives us confidence that stabilizing the environmental magnetic field for the two sensors on the sides of the experimental chamber will translate to the atoms within.

    If we engage our feedforward mechanism, i.e. only synthesize a 50-Hz out-of-phase AC signal, we see slightly narrowed distributions with $\sigma = 0.33$\,mG and 0.36\,mG; the signals are also strongly correlated at 95\%. A drastically different picture, however, is seen if we engage just the feedback mechanism: a bimodal distribution emerges instead. This behavior is consistent with the probability distribution of a sinewave, with an amplitude of about 0.3--0.4\,mG. The correlation between signals from the two fluxgate sensors remains high at 91\%. We note in passing that we operate under the assumption that quieting magnetic field interference at the position of the two sensors will also reduce the noise in the science chamber between them. This requires the magnetic field interference to be spatially uniform. Achieving consistent AC interference signals on both the MAIN and AUX sensors required the removal of several linear power supplies from above the optical table. Finally, engaging both the feedforward and feedback together reduces the field distribution width substantially, down to 93\,$\upmu$G and 105\,$\upmu$G. The signal correlation is also largely removed ($\rho = 36$\%) showing that the common-mode part of the magnetic field interference is now controlled.

    In Fig.~\ref{fig:histograms}~(b) we show magnetic field values on a per-shot basis. Here the position $B_i$ of a data point indicates the magnetic field averaged over the 100-ms window; the error bar denotes its standard deviation. Given that in our experiment the individual shots were separated by over 20\,s the scatter of the data points shows the slow shot-to-shot variation of the magnetic field offset that is best compensated with a feedback loop. The size of the error bars indicates standard deviation and is sensitive to the fast variation of the magnetic field within the short 100~ms time window. It is a good measure of the oscillatory AC signal that we combat with our synthesised feedforward signal and RMS noise.

    Again, without stabilizing the field we see both a scatter of the average field magnitudes (standard deviation of about 0.3\,mG) and AC field noise (also about 0.3\,mG). Turning on the feedforward reduces the RMS noise level to below  $100\,\upmu$G, but leaves the scatter untouched. On the other hand, engaging just the PID feedback loop obliterates the shot-to-shot variation in mean magnetic field: to $14\,\upmu$G on the MAIN and $52\,\upmu$G on the AUX sensor. Note that we use the MAIN sensor for the PID feedback loop, the AUX sensor acts merely as a spectator. Finally, engaging both feedback and feedforward enables us to keep both the slow drift and short term noise under $100\,\upmu$G.

 
 
 

\subsection{Cold-atom magnetometry}

\begin{figure}[b]
    \includegraphics[width=0.47\textwidth]{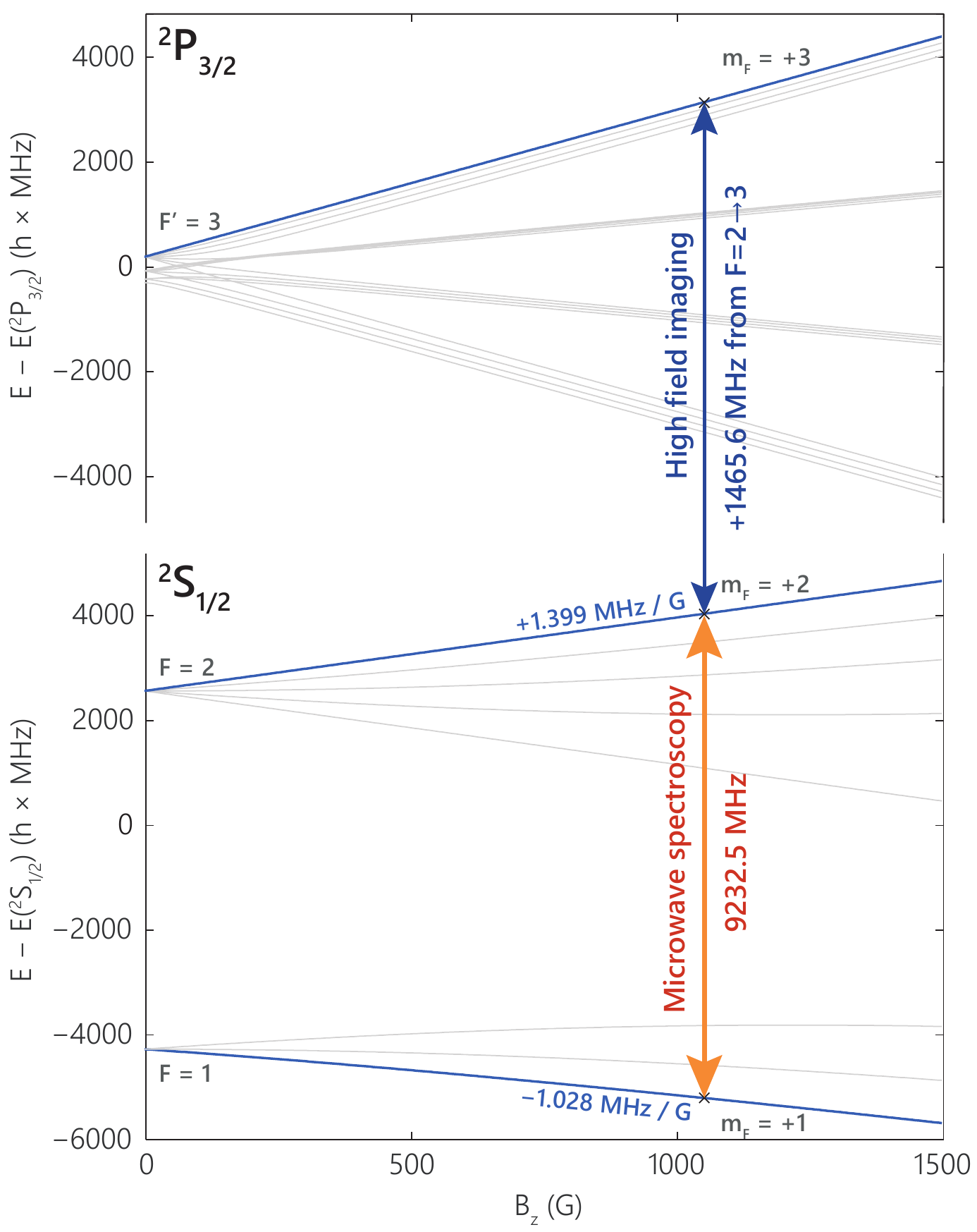}
    \caption{Microwave spectroscopy scheme for our characterization of magnetic field stability using ultracold atoms. First, we prepare ground-state Rb atoms in the $F=1, \,m_F=+1$ stretched state. Then, we interrogate the atoms with a microwave pulse tuned near a magnetic dipole transition to the $F=2, \,m_F=+2$ state. For a field $B=1050\,$G the differential Zeeman shift is $h\times 2.427\,{\rm MHz/G}$. Finally, the transferred atoms are imaged at the high magnetic field using the closed $^2$S$_{1/2}$~($F=2, \,m_F=+2$)$\leftrightarrow$$^2$P$_{3/2}$~($F'=3, \,m_F=+3$) transition at 780\,nm.}
    \label{fig:breitrabi}
\end{figure}

\begin{figure}[t]
    \includegraphics[width=0.47\textwidth]{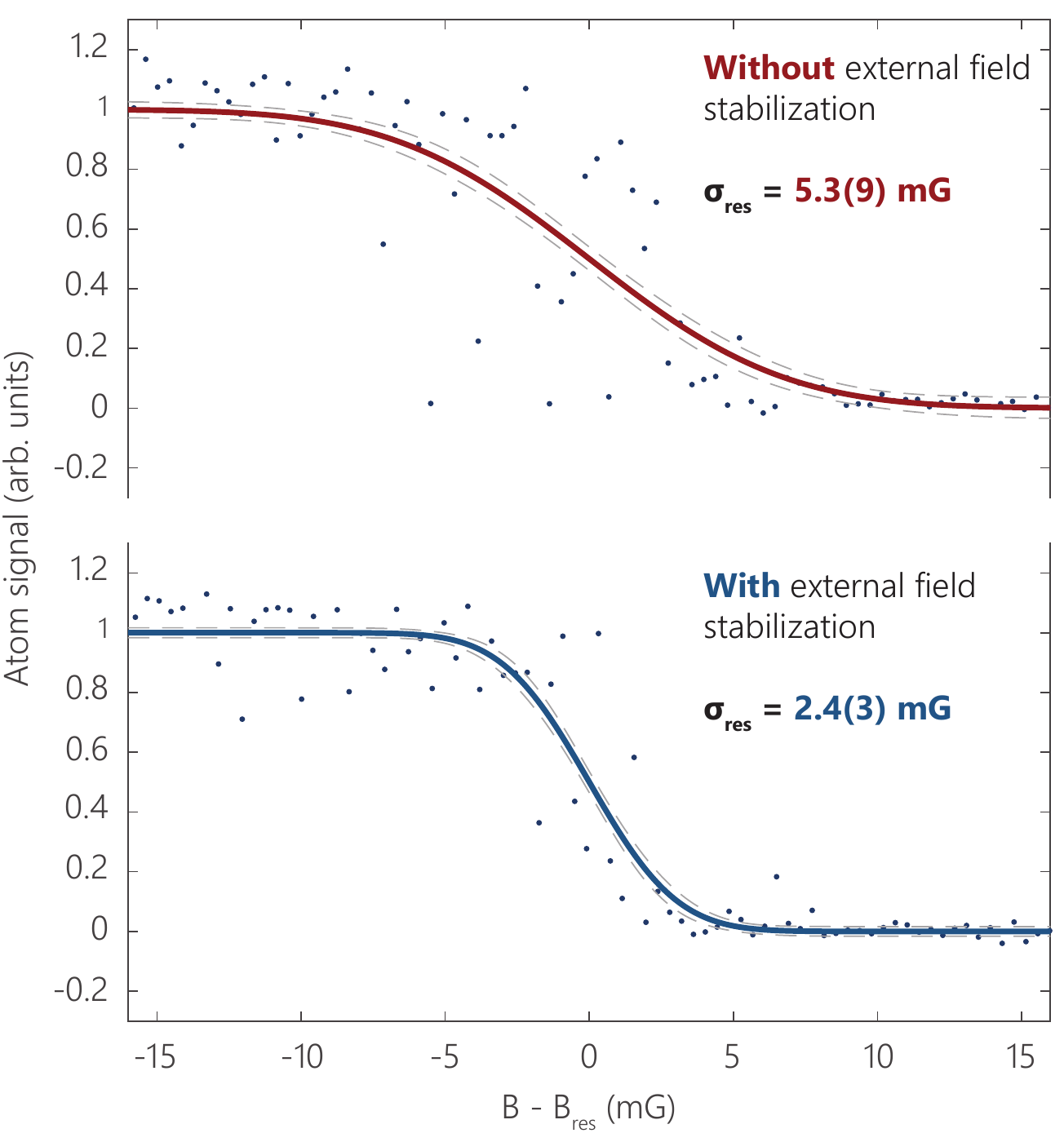}
    \caption{Impact of active stabilization on microwave spectra: a) no external magnetic field stabilization, b) full stabilization: both feedforward and feedback. Note that in both cases the Feshbach coil \emph{current} is actively stabilized; the difference lies in the active stabilization of the stray magnetic fields in the laboratory. In each measurement we sweep the microwave frequency over 50\,kHz, corresponding to about 20\,mG of magnetic field at 1050\,G. The sweep time is 20\,ms so that it covers one 50-Hz line cycle. The fitted functional form, Eq.~(\ref{eq:erf}), assumes a Gaussian shot-to-shot magnetic field distribution.}
    \label{fig:sweepspectra}
\end{figure}

    Our measurements of the magnetic field stability as experienced by our magnetic field sensors give us confidence that we can control the environmental magnetic fields to better than 1\,mG. This, however, is not the full picture -- the atoms in the trap are immersed in a combination of the magnetic field generated by our Feshbach coils and the environment. 

    To benchmark the magnetic field stability at the sample position we employ ultracold $^{87}$Rb atoms as magnetometers. The sample preparation procedure in our Rb-Sr machine is described in detail in Ref.~\cite{Pasquiou2013qdm}. To test our magnetic field stability we trap approximately $2\times 10^5$ ultracold Rb atoms in a spin mixture of all $m_F$ Zeeman substates of the $F = 1$ hyperfine manifold at a temperature of about 1\,$\mu$K in a 1064-nm dipole trap. We then purify the sample using a strong magnetic field gradient in order for the Stern-Gerlach force to remove the $m_F = \pm 1$ atoms from the trap. We ramp the magnetic field to 1319\,G and use an RF sweep from 744.7--744.8\,MHz to induce a rapid adiabatic transfer of the remaining $m_F = 0$ atoms to the $m_F = +1$ stretched state. We then ramp down the magnetic field to 1050\,G (311\,A coil current) over 100\,ms. Then, we engage our field stabilization PID and let it reach its final value over 650\,ms. This settling time is required because of a number of possible reasons, including eddy currents created while ramping the large Feshbach magnetic field as well as the settling of thermal effects generated in the coil. We found that a minimum wait time of 650 ms gave us the best performance in terms of the stability of the resonance position. We note that the field stabilization PID cannot compensate for those effects as it only stabilizes the external magnetic fields and not those generated by the Feshbach coil.

    Figure~\ref{fig:breitrabi} shows a Breit-Rabi diagram of the two Rb hyperfine ground states as a function of magnetic field. At a field of 1050\,G the differential Zeeman shift between the two stretched states is $2.427$\,MHz$/$G. Thus, by performing high resolution microwave spectroscopy on trapped Rb atoms we can directly measure the magnetic field fluctuations. After ensuring the magnetic field has stabilized to its final value we irradiate the atoms with up to 3\,W of 9232.5\,MHz microwave field using a log-periodic antenna (Aaronia AG HyperLOG 60180). We obtain our microwave signal using an HP 8672A synthesized signal generator and amplify it using a Kunhe Electronic KU PA 9001250-2A microwave amplifier. We use a linear 50\,kHz frequency sweep equivalent to about 20\,mG of change in magnetic field. The sweep time is 20\,ms, which corresponds to a single AC line cycle. We estimate the Rabi frequency to be on the order of 1 kHz based on a calibration of the rapid adiabatic passage process. This value of similar magnitude to the expected coupling strength of the 1313 G Feshbach resonance in RbSr~\cite{Ciamei2018trs}. The initial and final frequencies of the sweep are varied in lockstep; the final frequency is set close to the RF resonance. We observe the number of atoms transferred to the $F=2,\,m_F=+2$ stretched state using a high-field imaging system operating on the (semi) closed $^2$S$_{1/2}$~$F=2,\,m_F=+2$~$\leftrightarrow$~$^2$P$_{3/2}$~$F=3,\,m_F=+3$ transition at 780\,nm.

    Fig.~\ref{fig:sweepspectra} shows example lineshapes obtained using our microwave spectroscopy. If the RF resonance is well within the frequency sweep, then all atoms are transferred; conversely, if the resonance is well outside of the sweep, all of the atoms will remain in their initial state. The transition between the two regimes can be used to provide information about the magnetic field stability. If we assume that the magnetic field as experienced by the atoms has a normal distribution with a standard deviation $\sigma$, then the expected signal $S(B)$ is proportional to the cumulative distribution function of that distribution,
\begin{equation}
    S(B) = 
        S_{\rm bg} 
        + A \times \frac{1}{2}
        \left[
            1 - {\rm erf}\left(
            \frac{B - B_{\rm res}}{\sqrt{2} \sigma}
        \right)
        \right] 
        \label{eq:erf}
\end{equation}
    where $B_{\rm res}$ is the resonance position, while $S_{\rm bg}$ and $A$ are the signal background and amplitude. Here we also assumed, for simplicity, that the sweep range is much larger than the detuning from resonance. We stress that both the temporal and spatial (ie. gradient) magnetic field variation can contribute to~$\sigma$. Also, magnetic field instability is not the only broadening mechanism present here, so fitted values of $\sigma$ should be seen as an upper limit on the field fluctuations.

    The upper panel of Fig.~\ref{fig:sweepspectra} shows a microwave spectrum of Rb atoms when only the current stabilization in the coils is engaged, and the enviromental magnetic field is not actively stabilized, whether by our feedback or feedforward mechanisms. Fitting our simple lineshape model (Eq.~\ref{eq:erf}) to experimental data (upper panel of Fig.~\ref{fig:sweepspectra}) reveals an upper limit on the field standard deviation of $\sigma = 5.3(9)$\,mG. We also note significant scatter of data points around the resonance. Once we engage the field stabilization (lower panel of Fig.~\ref{fig:sweepspectra}), the scatter is significantly reduced and the fitted standard deviation drops down to $\sigma_{\rm res} = 2.4(3)$\,mG. This final result corresponds to a 2.3(3)\,ppm instability at 1050\,G. While we can identify several broadening mechanisms, they seem to be much weaker than the measured field instability. For example, we estimate that the residual field gradient created by a nearby ion pump is approximately 50 mG/cm, which over a cloud of about 10 micron in diameter would lead to a few tens of microgauss of broadening. The Fourier limit for our 20-ms pulse time is 2$\pi\times$50 Hz. Additionally, we have characterized our microwave spectroscopy setup by observing magnetic-field-insensitive clock transitions in Rb\cite{Harber2002} with a resolution of a few Hz, which gives us confidence that our resolution is not limited by the frequency source used, nor by optical trap shifts. 

\section{Conclusion}

    In conclusion we have demonstrated a two-stage magnetic field stabilization scheme that enables the creation of magnetic fields of over 1000\,G with ppm-level stability. We stabilize both the current flowing through the magnetic field coils as well as the environmental magnetic field. Our tests using microwave spectroscopy of Zeeman-split hyperfine levels of ultracold rubidium show an upper limit of magnetic field stability of 2.3(3)\,ppm or better at 1050\,G. This result should enable the magnetoassociation of ultracold molecules in cases where only ultranarrow magnetic Feshbach resonances at high fields are available. Thus, this work paves the way for experiments such as those in our laboratory, aimed at coherently producing ultracold samples of RbSr molecules\cite{Barbe2018ofr,Ciamei2018trs}. Precision Feshbach spectroscopy can also find use in fundamental physics: Efimov resonances\cite{Greene2017ufp,Ferlaino2011eru}, the tuning of $s$-wave interactions in quantum gases and in searches for the variation of fundamental constants\cite{Chin2006esf,Chin2010fri}. The schematics and codes for the magnetic field stabilization electronics can be found in an electronic repository~\cite{magnetic_field_schematics}.

\begin{acknowledgments}
    We thank Benjamin Pasquiou for support and advice during the early stages of this project. This work is part of the research programme “Atomic quantum simulators 2.0 -- taming the long-range interactions” with project number 680.92.18.05, which is funded by the Dutch Research Council (NWO). MB was partially funded by the National Science Centre of Poland OPUS project no. 2017/25/B/ST4/01486 and UWERTURA project no. 2019/32/U/ST2/00224. We gratefully acknowledge support from the Technology Centre of the University of Amsterdam.
\end{acknowledgments}

\section*{Author contributions}


    {\bf Mateusz Borkowski}: software (lead), writing - original draft (lead), conceptualization (equal), investigation (equal), resources (supporting).
    {\bf Lukas Reichs\"ollner}: resources (lead), conceptualization (lead), supervision (supporting), investigation (equal), software (supporting).
    {\bf Premjith Thekkeppatt}: resources (lead), data curation (lead), investigation (equal).
    {\bf Vincent Barb\'e}: Data curation (supporting), conceptualization (supporting), investigation (supporting).
    {\bf Tijs van Roon}: Resources (supporting), software (supporting).
    {\bf Klaasjan van Druten}: writing - original draft (supporting), resources (supporting), supervision (supporting).
    {\bf Florian Schreck}: Project administration (lead), resources (lead), supervision (lead), software (supporting).

\section*{Conflict of interest}
The authors have no conflicts to disclose.

\section*{Data availability}
The data that support the findings of this study are available from the corresponding author upon reasonable request. 
\bibliography{magfield}

\begin{thebibliography}{51}%
\makeatletter
\providecommand \@ifxundefined [1]{%
 \@ifx{#1\undefined}
}%
\providecommand \@ifnum [1]{%
 \ifnum #1\expandafter \@firstoftwo
 \else \expandafter \@secondoftwo
 \fi
}%
\providecommand \@ifx [1]{%
 \ifx #1\expandafter \@firstoftwo
 \else \expandafter \@secondoftwo
 \fi
}%
\providecommand \natexlab [1]{#1}%
\providecommand \enquote  [1]{``#1''}%
\providecommand \bibnamefont  [1]{#1}%
\providecommand \bibfnamefont [1]{#1}%
\providecommand \citenamefont [1]{#1}%
\providecommand \href@noop [0]{\@secondoftwo}%
\providecommand \href [0]{\begingroup \@sanitize@url \@href}%
\providecommand \@href[1]{\@@startlink{#1}\@@href}%
\providecommand \@@href[1]{\endgroup#1\@@endlink}%
\providecommand \@sanitize@url [0]{\catcode `\\12\catcode `\$12\catcode
  `\&12\catcode `\#12\catcode `\^12\catcode `\_12\catcode `\%12\relax}%
\providecommand \@@startlink[1]{}%
\providecommand \@@endlink[0]{}%
\providecommand \url  [0]{\begingroup\@sanitize@url \@url }%
\providecommand \@url [1]{\endgroup\@href {#1}{\urlprefix }}%
\providecommand \urlprefix  [0]{URL }%
\providecommand \Eprint [0]{\href }%
\providecommand \doibase [0]{https://doi.org/}%
\providecommand \selectlanguage [0]{\@gobble}%
\providecommand \bibinfo  [0]{\@secondoftwo}%
\providecommand \bibfield  [0]{\@secondoftwo}%
\providecommand \translation [1]{[#1]}%
\providecommand \BibitemOpen [0]{}%
\providecommand \bibitemStop [0]{}%
\providecommand \bibitemNoStop [0]{.\EOS\space}%
\providecommand \EOS [0]{\spacefactor3000\relax}%
\providecommand \BibitemShut  [1]{\csname bibitem#1\endcsname}%
\let\auto@bib@innerbib\@empty
\bibitem [{\citenamefont {Chin}\ \emph {et~al.}(2010)\citenamefont {Chin},
  \citenamefont {Grimm}, \citenamefont {Julienne},\ and\ \citenamefont
  {Tiesinga}}]{Chin2010fri}%
  \BibitemOpen
  \bibfield  {author} {\bibinfo {author} {\bibfnamefont {C.}~\bibnamefont
  {Chin}}, \bibinfo {author} {\bibfnamefont {R.}~\bibnamefont {Grimm}},
  \bibinfo {author} {\bibfnamefont {P.}~\bibnamefont {Julienne}},\ and\
  \bibinfo {author} {\bibfnamefont {E.}~\bibnamefont {Tiesinga}},\ }\bibfield
  {title} {\enquote {\bibinfo {title} {{Feshbach} resonances in ultracold
  gases},}\ }\href {https://doi.org/10.1103/revmodphys.82.1225} {\bibfield
  {journal} {\bibinfo  {journal} {Reviews of Modern Physics}\ }\textbf
  {\bibinfo {volume} {82}},\ \bibinfo {pages} {1225} (\bibinfo {year}
  {2010})}\BibitemShut {NoStop}%
\bibitem [{\citenamefont {Inouye}\ \emph {et~al.}(1998)\citenamefont {Inouye},
  \citenamefont {Andrews}, \citenamefont {Stenger}, \citenamefont {Mlesner},
  \citenamefont {Stamper-Kurn},\ and\ \citenamefont {Ketterle}}]{Inouye1998}%
  \BibitemOpen
  \bibfield  {author} {\bibinfo {author} {\bibfnamefont {S.}~\bibnamefont
  {Inouye}}, \bibinfo {author} {\bibfnamefont {M.~R.}\ \bibnamefont {Andrews}},
  \bibinfo {author} {\bibfnamefont {J.}~\bibnamefont {Stenger}}, \bibinfo
  {author} {\bibfnamefont {H.~J.}\ \bibnamefont {Mlesner}}, \bibinfo {author}
  {\bibfnamefont {D.~M.}\ \bibnamefont {Stamper-Kurn}},\ and\ \bibinfo {author}
  {\bibfnamefont {W.}~\bibnamefont {Ketterle}},\ }\bibfield  {title} {\enquote
  {\bibinfo {title} {{Observation of Feshbach resonances in a Bose–Einstein
  condensate}},}\ }\href {https://doi.org/10.1038/32354} {\bibfield  {journal}
  {\bibinfo  {journal} {Nature}\ }\textbf {\bibinfo {volume} {392}},\ \bibinfo
  {pages} {151} (\bibinfo {year} {1998})}\BibitemShut {NoStop}%
\bibitem [{\citenamefont {Bloch}, \citenamefont {Dalibard},\ and\ \citenamefont
  {Zwerger}(2008)}]{Bloch2008mbp}%
  \BibitemOpen
  \bibfield  {author} {\bibinfo {author} {\bibfnamefont {I.}~\bibnamefont
  {Bloch}}, \bibinfo {author} {\bibfnamefont {J.}~\bibnamefont {Dalibard}},\
  and\ \bibinfo {author} {\bibfnamefont {W.}~\bibnamefont {Zwerger}},\
  }\bibfield  {title} {\enquote {\bibinfo {title} {Many-body physics with
  ultracold gases},}\ }\href {https://doi.org/10.1103/RevModPhys.80.885}
  {\bibfield  {journal} {\bibinfo  {journal} {Rev. Mod. Phys.}\ }\textbf
  {\bibinfo {volume} {80}},\ \bibinfo {pages} {885} (\bibinfo {year}
  {2008})}\BibitemShut {NoStop}%
\bibitem [{\citenamefont {Cornish}\ \emph {et~al.}(2000)\citenamefont
  {Cornish}, \citenamefont {Claussen}, \citenamefont {Roberts}, \citenamefont
  {Cornell},\ and\ \citenamefont {Wieman}}]{Cornish2000}%
  \BibitemOpen
  \bibfield  {author} {\bibinfo {author} {\bibfnamefont {S.~L.}\ \bibnamefont
  {Cornish}}, \bibinfo {author} {\bibfnamefont {N.~R.}\ \bibnamefont
  {Claussen}}, \bibinfo {author} {\bibfnamefont {J.~L.}\ \bibnamefont
  {Roberts}}, \bibinfo {author} {\bibfnamefont {E.~A.}\ \bibnamefont
  {Cornell}},\ and\ \bibinfo {author} {\bibfnamefont {C.~E.}\ \bibnamefont
  {Wieman}},\ }\bibfield  {title} {\enquote {\bibinfo {title} {{Stable
  $^{85}$Rb Bose-Einstein Condensates with Widely Tunable Interactions}},}\
  }\href {https://doi.org/10.1103/PhysRevLett.85.1795} {\bibfield  {journal}
  {\bibinfo  {journal} {Physical Review Letters}\ }\textbf {\bibinfo {volume}
  {85}},\ \bibinfo {pages} {1795} (\bibinfo {year} {2000})}\BibitemShut
  {NoStop}%
\bibitem [{\citenamefont {Roberts}\ \emph {et~al.}(2001)\citenamefont
  {Roberts}, \citenamefont {Claussen}, \citenamefont {Cornish}, \citenamefont
  {Donley}, \citenamefont {Cornell},\ and\ \citenamefont
  {Wieman}}]{Roberts2001}%
  \BibitemOpen
  \bibfield  {author} {\bibinfo {author} {\bibfnamefont {J.~L.}\ \bibnamefont
  {Roberts}}, \bibinfo {author} {\bibfnamefont {N.~R.}\ \bibnamefont
  {Claussen}}, \bibinfo {author} {\bibfnamefont {S.~L.}\ \bibnamefont
  {Cornish}}, \bibinfo {author} {\bibfnamefont {E.~A.}\ \bibnamefont {Donley}},
  \bibinfo {author} {\bibfnamefont {E.~A.}\ \bibnamefont {Cornell}},\ and\
  \bibinfo {author} {\bibfnamefont {C.~E.}\ \bibnamefont {Wieman}},\ }\bibfield
   {title} {\enquote {\bibinfo {title} {{Controlled Collapse of a Bose-Einstein
  Condensate}},}\ }\href {https://doi.org/10.1103/PhysRevLett.86.4211}
  {\bibfield  {journal} {\bibinfo  {journal} {Physical Review Letters}\
  }\textbf {\bibinfo {volume} {86}},\ \bibinfo {pages} {4211} (\bibinfo {year}
  {2001})}\BibitemShut {NoStop}%
\bibitem [{\citenamefont {Donley}\ \emph {et~al.}(2001)\citenamefont {Donley},
  \citenamefont {Claussen}, \citenamefont {Cornish}, \citenamefont {Roberts},
  \citenamefont {Cornell},\ and\ \citenamefont {Wieman}}]{Donley2001}%
  \BibitemOpen
  \bibfield  {author} {\bibinfo {author} {\bibfnamefont {E.~A.}\ \bibnamefont
  {Donley}}, \bibinfo {author} {\bibfnamefont {N.~R.}\ \bibnamefont
  {Claussen}}, \bibinfo {author} {\bibfnamefont {S.~L.}\ \bibnamefont
  {Cornish}}, \bibinfo {author} {\bibfnamefont {J.~L.}\ \bibnamefont
  {Roberts}}, \bibinfo {author} {\bibfnamefont {E.~A.}\ \bibnamefont
  {Cornell}},\ and\ \bibinfo {author} {\bibfnamefont {C.~E.}\ \bibnamefont
  {Wieman}},\ }\bibfield  {title} {\enquote {\bibinfo {title} {{Dynamics of
  collapsing and exploding Bose–Einstein condensates}},}\ }\href
  {https://doi.org/10.1038/35085500} {\bibfield  {journal} {\bibinfo  {journal}
  {Nature}\ }\textbf {\bibinfo {volume} {412}},\ \bibinfo {pages} {295}
  (\bibinfo {year} {2001})}\BibitemShut {NoStop}%
\bibitem [{\citenamefont {Parish}(2014)}]{Parish2014bcs}%
  \BibitemOpen
  \bibfield  {author} {\bibinfo {author} {\bibfnamefont {M.~M.}\ \bibnamefont
  {Parish}},\ }\bibfield  {title} {\enquote {\bibinfo {title} {The {BCS}--{BEC}
  crossover},}\ }in\ \href {https://doi.org/10.1142/9781783264766_0009} {\emph
  {\bibinfo {booktitle} {Quantum Gas experiments, Exploring Many-Body
  States}}}\ (\bibinfo  {publisher} {{Imperial} {College} {Press}},\ \bibinfo
  {year} {2014})\ p.\ \bibinfo {pages} {179}\BibitemShut {NoStop}%
\bibitem [{\citenamefont {Jochim}\ \emph {et~al.}(2003)\citenamefont {Jochim},
  \citenamefont {Bartenstein}, \citenamefont {Altmeyer}, \citenamefont {Hendl},
  \citenamefont {Riedl}, \citenamefont {Chin}, \citenamefont {{Hecker
  Denschtag}},\ and\ \citenamefont {Grimm}}]{Jochim2003}%
  \BibitemOpen
  \bibfield  {author} {\bibinfo {author} {\bibfnamefont {S.}~\bibnamefont
  {Jochim}}, \bibinfo {author} {\bibfnamefont {M.}~\bibnamefont {Bartenstein}},
  \bibinfo {author} {\bibfnamefont {A.}~\bibnamefont {Altmeyer}}, \bibinfo
  {author} {\bibfnamefont {G.}~\bibnamefont {Hendl}}, \bibinfo {author}
  {\bibfnamefont {S.}~\bibnamefont {Riedl}}, \bibinfo {author} {\bibfnamefont
  {C.}~\bibnamefont {Chin}}, \bibinfo {author} {\bibfnamefont {J.}~\bibnamefont
  {{Hecker Denschtag}}},\ and\ \bibinfo {author} {\bibfnamefont
  {R.}~\bibnamefont {Grimm}},\ }\bibfield  {title} {\enquote {\bibinfo {title}
  {{Bose-Einstein Condensation of Molecules}},}\ }\href
  {https://doi.org/10.1126/SCIENCE.1093280} {\bibfield  {journal} {\bibinfo
  {journal} {Science}\ }\textbf {\bibinfo {volume} {302}},\ \bibinfo {pages}
  {2101} (\bibinfo {year} {2003})}\BibitemShut {NoStop}%
\bibitem [{\citenamefont {K{\"o}hler}, \citenamefont {G\'oral},\ and\
  \citenamefont {Julienne}(2006)}]{Kohler2006poc}%
  \BibitemOpen
  \bibfield  {author} {\bibinfo {author} {\bibfnamefont {T.}~\bibnamefont
  {K{\"o}hler}}, \bibinfo {author} {\bibfnamefont {K.}~\bibnamefont
  {G\'oral}},\ and\ \bibinfo {author} {\bibfnamefont {P.~S.}\ \bibnamefont
  {Julienne}},\ }\bibfield  {title} {\enquote {\bibinfo {title} {Production of
  cold molecules via magnetically tunable {Feshbach} resonances},}\ }\href
  {https://doi.org/10.1103/revmodphys.78.1311} {\bibfield  {journal} {\bibinfo
  {journal} {Reviews of Modern Physics}\ }\textbf {\bibinfo {volume} {78}},\
  \bibinfo {eid} {1311} (\bibinfo {year} {2006})}\BibitemShut {NoStop}%
\bibitem [{\citenamefont {Weckesser}\ \emph {et~al.}(2021)\citenamefont
  {Weckesser}, \citenamefont {Thielemann}, \citenamefont {Wiater},
  \citenamefont {Wojciechowska}, \citenamefont {Karpa}, \citenamefont
  {Jachymski}, \citenamefont {Tomza}, \citenamefont {Walker},\ and\
  \citenamefont {Schaetz}}]{Weckesser2021ofr}%
  \BibitemOpen
  \bibfield  {author} {\bibinfo {author} {\bibfnamefont {P.}~\bibnamefont
  {Weckesser}}, \bibinfo {author} {\bibfnamefont {F.}~\bibnamefont
  {Thielemann}}, \bibinfo {author} {\bibfnamefont {D.}~\bibnamefont {Wiater}},
  \bibinfo {author} {\bibfnamefont {A.}~\bibnamefont {Wojciechowska}}, \bibinfo
  {author} {\bibfnamefont {L.}~\bibnamefont {Karpa}}, \bibinfo {author}
  {\bibfnamefont {K.}~\bibnamefont {Jachymski}}, \bibinfo {author}
  {\bibfnamefont {M.}~\bibnamefont {Tomza}}, \bibinfo {author} {\bibfnamefont
  {T.}~\bibnamefont {Walker}},\ and\ \bibinfo {author} {\bibfnamefont
  {T.}~\bibnamefont {Schaetz}},\ }\bibfield  {title} {\enquote {\bibinfo
  {title} {Observation of {Feshbach} resonances between a single ion and
  ultracold atoms},}\ }\href {https://doi.org/10.1038/s41586-021-04112-y}
  {\bibfield  {journal} {\bibinfo  {journal} {Nature}\ }\textbf {\bibinfo
  {volume} {600}},\ \bibinfo {pages} {429} (\bibinfo {year}
  {2021})}\BibitemShut {NoStop}%
\bibitem [{\citenamefont {Greene}, \citenamefont {Giannakeas},\ and\
  \citenamefont {P\'erez-R\'{\i}os}(2017)}]{Greene2017ufp}%
  \BibitemOpen
  \bibfield  {author} {\bibinfo {author} {\bibfnamefont {C.~H.}\ \bibnamefont
  {Greene}}, \bibinfo {author} {\bibfnamefont {P.}~\bibnamefont {Giannakeas}},\
  and\ \bibinfo {author} {\bibfnamefont {J.}~\bibnamefont
  {P\'erez-R\'{\i}os}},\ }\bibfield  {title} {\enquote {\bibinfo {title}
  {Universal few-body physics and cluster formation},}\ }\href
  {https://doi.org/10.1103/RevModPhys.89.035006} {\bibfield  {journal}
  {\bibinfo  {journal} {Rev. Mod. Phys.}\ }\textbf {\bibinfo {volume} {89}},\
  \bibinfo {pages} {035006} (\bibinfo {year} {2017})}\BibitemShut {NoStop}%
\bibitem [{\citenamefont {Ferlaino}\ \emph {et~al.}(2011)\citenamefont
  {Ferlaino}, \citenamefont {Zenesini}, \citenamefont {Berninger},
  \citenamefont {Huang}, \citenamefont {N{\"{a}}gerl},\ and\ \citenamefont
  {Grimm}}]{Ferlaino2011eru}%
  \BibitemOpen
  \bibfield  {author} {\bibinfo {author} {\bibfnamefont {F.}~\bibnamefont
  {Ferlaino}}, \bibinfo {author} {\bibfnamefont {A.}~\bibnamefont {Zenesini}},
  \bibinfo {author} {\bibfnamefont {M.}~\bibnamefont {Berninger}}, \bibinfo
  {author} {\bibfnamefont {B.}~\bibnamefont {Huang}}, \bibinfo {author}
  {\bibfnamefont {H.-C.}\ \bibnamefont {N{\"{a}}gerl}},\ and\ \bibinfo {author}
  {\bibfnamefont {R.}~\bibnamefont {Grimm}},\ }\bibfield  {title} {\enquote
  {\bibinfo {title} {{Efimov Resonances in Ultracold Quantum Gases}},}\ }\href
  {https://doi.org/10.1007/s00601-011-0260-7} {\bibfield  {journal} {\bibinfo
  {journal} {Few-Body Systems}\ }\textbf {\bibinfo {volume} {51}},\ \bibinfo
  {pages} {113--133} (\bibinfo {year} {2011})}\BibitemShut {NoStop}%
\bibitem [{\citenamefont {Chin}\ and\ \citenamefont
  {Flambaum}(2006)}]{Chin2006esf}%
  \BibitemOpen
  \bibfield  {author} {\bibinfo {author} {\bibfnamefont {C.}~\bibnamefont
  {Chin}}\ and\ \bibinfo {author} {\bibfnamefont {V.~V.}\ \bibnamefont
  {Flambaum}},\ }\bibfield  {title} {\enquote {\bibinfo {title} {{Enhanced
  Sensitivity to Fundamental Constants In Ultracold Atomic and Molecular
  Systems near Feshbach Resonances}},}\ }\href
  {https://doi.org/10.1103/PhysRevLett.96.230801} {\bibfield  {journal}
  {\bibinfo  {journal} {Physical Review Letters}\ }\textbf {\bibinfo {volume}
  {96}},\ \bibinfo {pages} {230801} (\bibinfo {year} {2006})}\BibitemShut
  {NoStop}%
\bibitem [{\citenamefont {Barb{\'{e}}}\ \emph {et~al.}(2018)\citenamefont
  {Barb{\'{e}}}, \citenamefont {Ciamei}, \citenamefont {Pasquiou},
  \citenamefont {Reichsöllner}, \citenamefont {Schreck}, \citenamefont
  {{\.{Z}}uchowski},\ and\ \citenamefont {Hutson}}]{Barbe2018ofr}%
  \BibitemOpen
  \bibfield  {author} {\bibinfo {author} {\bibfnamefont {V.}~\bibnamefont
  {Barb{\'{e}}}}, \bibinfo {author} {\bibfnamefont {A.}~\bibnamefont {Ciamei}},
  \bibinfo {author} {\bibfnamefont {B.}~\bibnamefont {Pasquiou}}, \bibinfo
  {author} {\bibfnamefont {L.}~\bibnamefont {Reichsöllner}}, \bibinfo {author}
  {\bibfnamefont {F.}~\bibnamefont {Schreck}}, \bibinfo {author} {\bibfnamefont
  {P.~S.}\ \bibnamefont {{\.{Z}}uchowski}},\ and\ \bibinfo {author}
  {\bibfnamefont {J.~M.}\ \bibnamefont {Hutson}},\ }\bibfield  {title}
  {\enquote {\bibinfo {title} {Observation of {Feshbach} resonances between
  alkali and closed-shell atoms},}\ }\href
  {https://doi.org/10.1038/s41567-018-0169-x} {\bibfield  {journal} {\bibinfo
  {journal} {Nature Physics}\ }\textbf {\bibinfo {volume} {14}},\ \bibinfo
  {pages} {881} (\bibinfo {year} {2018})}\BibitemShut {NoStop}%
\bibitem [{\citenamefont {Ciamei}\ \emph {et~al.}(2018)\citenamefont {Ciamei},
  \citenamefont {Szczepkowski}, \citenamefont {Bayerle}, \citenamefont
  {Barbé}, \citenamefont {Reichsöllner}, \citenamefont {Tzanova},
  \citenamefont {Chen}, \citenamefont {Pasquiou}, \citenamefont {Grochola},
  \citenamefont {Kowalczyk}, \citenamefont {Jastrzebski},\ and\ \citenamefont
  {Schreck}}]{Ciamei2018trs}%
  \BibitemOpen
  \bibfield  {author} {\bibinfo {author} {\bibfnamefont {A.}~\bibnamefont
  {Ciamei}}, \bibinfo {author} {\bibfnamefont {J.}~\bibnamefont
  {Szczepkowski}}, \bibinfo {author} {\bibfnamefont {A.}~\bibnamefont
  {Bayerle}}, \bibinfo {author} {\bibfnamefont {V.}~\bibnamefont {Barbé}},
  \bibinfo {author} {\bibfnamefont {L.}~\bibnamefont {Reichsöllner}}, \bibinfo
  {author} {\bibfnamefont {S.~M.}\ \bibnamefont {Tzanova}}, \bibinfo {author}
  {\bibfnamefont {C.-C.}\ \bibnamefont {Chen}}, \bibinfo {author}
  {\bibfnamefont {B.}~\bibnamefont {Pasquiou}}, \bibinfo {author}
  {\bibfnamefont {A.}~\bibnamefont {Grochola}}, \bibinfo {author}
  {\bibfnamefont {P.}~\bibnamefont {Kowalczyk}}, \bibinfo {author}
  {\bibfnamefont {W.}~\bibnamefont {Jastrzebski}},\ and\ \bibinfo {author}
  {\bibfnamefont {F.}~\bibnamefont {Schreck}},\ }\bibfield  {title} {\enquote
  {\bibinfo {title} {The {R}b{S}r $^{2}{\Sigma}^{+}$ ground state investigated
  via spectroscopy of hot and ultracold molecules},}\ }\href
  {https://doi.org/https://doi.org/10.1039/c8cp03919d} {\bibfield  {journal}
  {\bibinfo  {journal} {Phys. Chem. Chem. Phys.}\ }\textbf {\bibinfo {volume}
  {20}},\ \bibinfo {pages} {26221} (\bibinfo {year} {2018})}\BibitemShut
  {NoStop}%
\bibitem [{\citenamefont {Regal}\ \emph {et~al.}(2003)\citenamefont {Regal},
  \citenamefont {Ticknor}, \citenamefont {Bohn},\ and\ \citenamefont
  {Jin}}]{Regal2003}%
  \BibitemOpen
  \bibfield  {author} {\bibinfo {author} {\bibfnamefont {C.~A.}\ \bibnamefont
  {Regal}}, \bibinfo {author} {\bibfnamefont {C.}~\bibnamefont {Ticknor}},
  \bibinfo {author} {\bibfnamefont {J.~L.}\ \bibnamefont {Bohn}},\ and\
  \bibinfo {author} {\bibfnamefont {D.~S.}\ \bibnamefont {Jin}},\ }\bibfield
  {title} {\enquote {\bibinfo {title} {{Creation of ultracold molecules from a
  Fermi gas of atoms}},}\ }\href {https://doi.org/10.1038/nature01738}
  {\bibfield  {journal} {\bibinfo  {journal} {Nature}\ }\textbf {\bibinfo
  {volume} {424}},\ \bibinfo {pages} {47} (\bibinfo {year} {2003})}\BibitemShut
  {NoStop}%
\bibitem [{\citenamefont {Herbig}\ \emph {et~al.}(2003)\citenamefont {Herbig},
  \citenamefont {Kraemer}, \citenamefont {Mark}, \citenamefont {Weber},
  \citenamefont {Chin}, \citenamefont {N{\"{a}}gerl},\ and\ \citenamefont
  {Grimm}}]{Herbig2003}%
  \BibitemOpen
  \bibfield  {author} {\bibinfo {author} {\bibfnamefont {J.}~\bibnamefont
  {Herbig}}, \bibinfo {author} {\bibfnamefont {T.}~\bibnamefont {Kraemer}},
  \bibinfo {author} {\bibfnamefont {M.}~\bibnamefont {Mark}}, \bibinfo {author}
  {\bibfnamefont {T.}~\bibnamefont {Weber}}, \bibinfo {author} {\bibfnamefont
  {C.}~\bibnamefont {Chin}}, \bibinfo {author} {\bibfnamefont {H.~C.}\
  \bibnamefont {N{\"{a}}gerl}},\ and\ \bibinfo {author} {\bibfnamefont
  {R.}~\bibnamefont {Grimm}},\ }\bibfield  {title} {\enquote {\bibinfo {title}
  {{Preparation of a pure molecular quantum gas}},}\ }\href
  {https://doi.org/10.1126/SCIENCE.1088876} {\bibfield  {journal} {\bibinfo
  {journal} {Science}\ }\textbf {\bibinfo {volume} {301}},\ \bibinfo {pages}
  {1510} (\bibinfo {year} {2003})}\BibitemShut {NoStop}%
\bibitem [{\citenamefont {Ni}\ \emph {et~al.}(2008)\citenamefont {Ni},
  \citenamefont {Ospelkaus}, \citenamefont {{De Miranda}}, \citenamefont
  {Pe'er}, \citenamefont {Neyenhuis}, \citenamefont {Zirbel}, \citenamefont
  {Kotochigova}, \citenamefont {Julienne}, \citenamefont {Jin},\ and\
  \citenamefont {Ye}}]{Ni2008}%
  \BibitemOpen
  \bibfield  {author} {\bibinfo {author} {\bibfnamefont {K.~K.}\ \bibnamefont
  {Ni}}, \bibinfo {author} {\bibfnamefont {S.}~\bibnamefont {Ospelkaus}},
  \bibinfo {author} {\bibfnamefont {M.~H.}\ \bibnamefont {{De Miranda}}},
  \bibinfo {author} {\bibfnamefont {A.}~\bibnamefont {Pe'er}}, \bibinfo
  {author} {\bibfnamefont {B.}~\bibnamefont {Neyenhuis}}, \bibinfo {author}
  {\bibfnamefont {J.~J.}\ \bibnamefont {Zirbel}}, \bibinfo {author}
  {\bibfnamefont {S.}~\bibnamefont {Kotochigova}}, \bibinfo {author}
  {\bibfnamefont {P.~S.}\ \bibnamefont {Julienne}}, \bibinfo {author}
  {\bibfnamefont {D.~S.}\ \bibnamefont {Jin}},\ and\ \bibinfo {author}
  {\bibfnamefont {J.}~\bibnamefont {Ye}},\ }\bibfield  {title} {\enquote
  {\bibinfo {title} {{A high phase-space-density gas of polar molecules}},}\
  }\href {https://doi.org/10.1126/SCIENCE.1163861} {\bibfield  {journal}
  {\bibinfo  {journal} {Science}\ }\textbf {\bibinfo {volume} {322}},\ \bibinfo
  {pages} {231} (\bibinfo {year} {2008})}\BibitemShut {NoStop}%
\bibitem [{\citenamefont {Danzl}\ \emph {et~al.}(2010)\citenamefont {Danzl},
  \citenamefont {Mark}, \citenamefont {Haller}, \citenamefont {Gustavsson},
  \citenamefont {Hart}, \citenamefont {Aldegunde}, \citenamefont {Hutson},\
  and\ \citenamefont {N{\"{a}}gerl}}]{Danzl2010}%
  \BibitemOpen
  \bibfield  {author} {\bibinfo {author} {\bibfnamefont {J.~G.}\ \bibnamefont
  {Danzl}}, \bibinfo {author} {\bibfnamefont {M.~J.}\ \bibnamefont {Mark}},
  \bibinfo {author} {\bibfnamefont {E.}~\bibnamefont {Haller}}, \bibinfo
  {author} {\bibfnamefont {M.}~\bibnamefont {Gustavsson}}, \bibinfo {author}
  {\bibfnamefont {R.}~\bibnamefont {Hart}}, \bibinfo {author} {\bibfnamefont
  {J.}~\bibnamefont {Aldegunde}}, \bibinfo {author} {\bibfnamefont {J.~M.}\
  \bibnamefont {Hutson}},\ and\ \bibinfo {author} {\bibfnamefont {H.-C.}\
  \bibnamefont {N{\"{a}}gerl}},\ }\bibfield  {title} {\enquote {\bibinfo
  {title} {{An ultracold high-density sample of rovibronic ground-state
  molecules in an optical lattice}},}\ }\href
  {https://doi.org/10.1038/nphys1533} {\bibfield  {journal} {\bibinfo
  {journal} {Nature Physics}\ }\textbf {\bibinfo {volume} {6}},\ \bibinfo
  {pages} {265} (\bibinfo {year} {2010})}\BibitemShut {NoStop}%
\bibitem [{\citenamefont {Takekoshi}\ \emph {et~al.}(2014)\citenamefont
  {Takekoshi}, \citenamefont {Reichs{\"{o}}llner}, \citenamefont {Schindewolf},
  \citenamefont {Hutson}, \citenamefont {{Le Sueur}}, \citenamefont {Dulieu},
  \citenamefont {Ferlaino}, \citenamefont {Grimm},\ and\ \citenamefont
  {N{\"{a}}gerl}}]{Takekoshi2014a}%
  \BibitemOpen
  \bibfield  {author} {\bibinfo {author} {\bibfnamefont {T.}~\bibnamefont
  {Takekoshi}}, \bibinfo {author} {\bibfnamefont {L.}~\bibnamefont
  {Reichs{\"{o}}llner}}, \bibinfo {author} {\bibfnamefont {A.}~\bibnamefont
  {Schindewolf}}, \bibinfo {author} {\bibfnamefont {J.~M.}\ \bibnamefont
  {Hutson}}, \bibinfo {author} {\bibfnamefont {C.~R.}\ \bibnamefont {{Le
  Sueur}}}, \bibinfo {author} {\bibfnamefont {O.}~\bibnamefont {Dulieu}},
  \bibinfo {author} {\bibfnamefont {F.}~\bibnamefont {Ferlaino}}, \bibinfo
  {author} {\bibfnamefont {R.}~\bibnamefont {Grimm}},\ and\ \bibinfo {author}
  {\bibfnamefont {H.~C.}\ \bibnamefont {N{\"{a}}gerl}},\ }\bibfield  {title}
  {\enquote {\bibinfo {title} {{Ultracold dense samples of dipolar RbCs
  molecules in the rovibrational and hyperfine ground state}},}\ }\href
  {https://doi.org/10.1103/PHYSREVLETT.113.205301} {\bibfield  {journal}
  {\bibinfo  {journal} {Physical Review Letters}\ }\textbf {\bibinfo {volume}
  {113}},\ \bibinfo {pages} {205301} (\bibinfo {year} {2014})}\BibitemShut
  {NoStop}%
\bibitem [{\citenamefont {Moses}\ \emph {et~al.}(2015)\citenamefont {Moses},
  \citenamefont {Covey}, \citenamefont {Miecnikowski}, \citenamefont {Yan},
  \citenamefont {Gadway}, \citenamefont {Ye},\ and\ \citenamefont
  {Jin}}]{Moses2015}%
  \BibitemOpen
  \bibfield  {author} {\bibinfo {author} {\bibfnamefont {S.~A.}\ \bibnamefont
  {Moses}}, \bibinfo {author} {\bibfnamefont {J.~P.}\ \bibnamefont {Covey}},
  \bibinfo {author} {\bibfnamefont {M.~T.}\ \bibnamefont {Miecnikowski}},
  \bibinfo {author} {\bibfnamefont {B.}~\bibnamefont {Yan}}, \bibinfo {author}
  {\bibfnamefont {B.}~\bibnamefont {Gadway}}, \bibinfo {author} {\bibfnamefont
  {J.}~\bibnamefont {Ye}},\ and\ \bibinfo {author} {\bibfnamefont {D.~S.}\
  \bibnamefont {Jin}},\ }\bibfield  {title} {\enquote {\bibinfo {title}
  {{Creation of a low-entropy quantum gas of polar molecules in an optical
  lattice}},}\ }\href {https://doi.org/10.1126/SCIENCE.AAC6400} {\bibfield
  {journal} {\bibinfo  {journal} {Science}\ }\textbf {\bibinfo {volume}
  {350}},\ \bibinfo {pages} {659} (\bibinfo {year} {2015})}\BibitemShut
  {NoStop}%
\bibitem [{\citenamefont {Reichs{\"{o}}llner}\ \emph
  {et~al.}(2017)\citenamefont {Reichs{\"{o}}llner}, \citenamefont
  {Schindewolf}, \citenamefont {Takekoshi}, \citenamefont {Grimm},\ and\
  \citenamefont {N{\"{a}}gerl}}]{Reichsollner2017}%
  \BibitemOpen
  \bibfield  {author} {\bibinfo {author} {\bibfnamefont {L.}~\bibnamefont
  {Reichs{\"{o}}llner}}, \bibinfo {author} {\bibfnamefont {A.}~\bibnamefont
  {Schindewolf}}, \bibinfo {author} {\bibfnamefont {T.}~\bibnamefont
  {Takekoshi}}, \bibinfo {author} {\bibfnamefont {R.}~\bibnamefont {Grimm}},\
  and\ \bibinfo {author} {\bibfnamefont {H.~C.}\ \bibnamefont {N{\"{a}}gerl}},\
  }\bibfield  {title} {\enquote {\bibinfo {title} {{Quantum Engineering of a
  Low-Entropy Gas of Heteronuclear Bosonic Molecules in an Optical Lattice}},}\
  }\href {https://doi.org/10.1103/PHYSREVLETT.118.073201} {\bibfield  {journal}
  {\bibinfo  {journal} {Physical Review Letters}\ }\textbf {\bibinfo {volume}
  {118}},\ \bibinfo {pages} {073201} (\bibinfo {year} {2017})}\BibitemShut
  {NoStop}%
\bibitem [{\citenamefont {Voges}\ \emph {et~al.}(2020)\citenamefont {Voges},
  \citenamefont {Gersema}, \citenamefont {Hartmann}, \citenamefont {Schulze},
  \citenamefont {Zenesini},\ and\ \citenamefont {Ospelkaus}}]{Voges2020}%
  \BibitemOpen
  \bibfield  {author} {\bibinfo {author} {\bibfnamefont {K.~K.}\ \bibnamefont
  {Voges}}, \bibinfo {author} {\bibfnamefont {P.}~\bibnamefont {Gersema}},
  \bibinfo {author} {\bibfnamefont {T.}~\bibnamefont {Hartmann}}, \bibinfo
  {author} {\bibfnamefont {T.~A.}\ \bibnamefont {Schulze}}, \bibinfo {author}
  {\bibfnamefont {A.}~\bibnamefont {Zenesini}},\ and\ \bibinfo {author}
  {\bibfnamefont {S.}~\bibnamefont {Ospelkaus}},\ }\bibfield  {title} {\enquote
  {\bibinfo {title} {{Formation of ultracold weakly bound dimers of bosonic
  $^{23}$Na$^{39}$K}},}\ }\href {https://doi.org/10.1103/PHYSREVA.101.042704}
  {\bibfield  {journal} {\bibinfo  {journal} {Physical Review A}\ }\textbf
  {\bibinfo {volume} {101}},\ \bibinfo {pages} {042704} (\bibinfo {year}
  {2020})}\BibitemShut {NoStop}%
\bibitem [{\citenamefont {Vitanov}\ \emph {et~al.}(2017)\citenamefont
  {Vitanov}, \citenamefont {Rangelov}, \citenamefont {Shore},\ and\
  \citenamefont {Bergmann}}]{Vitanov2017}%
  \BibitemOpen
  \bibfield  {author} {\bibinfo {author} {\bibfnamefont {N.~V.}\ \bibnamefont
  {Vitanov}}, \bibinfo {author} {\bibfnamefont {A.~A.}\ \bibnamefont
  {Rangelov}}, \bibinfo {author} {\bibfnamefont {B.~W.}\ \bibnamefont
  {Shore}},\ and\ \bibinfo {author} {\bibfnamefont {K.}~\bibnamefont
  {Bergmann}},\ }\bibfield  {title} {\enquote {\bibinfo {title} {{Stimulated
  Raman adiabatic passage in physics, chemistry, and beyond}},}\ }\href
  {https://doi.org/10.1103/RevModPhys.89.015006} {\bibfield  {journal}
  {\bibinfo  {journal} {Reviews of Modern Physics}\ }\textbf {\bibinfo {volume}
  {89}},\ \bibinfo {pages} {1} (\bibinfo {year} {2017})}\BibitemShut {NoStop}%
\bibitem [{\citenamefont {DeMille}(2002)}]{Demille2002}%
  \BibitemOpen
  \bibfield  {author} {\bibinfo {author} {\bibfnamefont {D.}~\bibnamefont
  {DeMille}},\ }\bibfield  {title} {\enquote {\bibinfo {title} {{Quantum
  Computation with Trapped Polar Molecules}},}\ }\href
  {https://doi.org/10.1103/PhysRevLett.88.067901} {\bibfield  {journal}
  {\bibinfo  {journal} {Physical Review Letters}\ }\textbf {\bibinfo {volume}
  {88}},\ \bibinfo {pages} {067901} (\bibinfo {year} {2002})}\BibitemShut
  {NoStop}%
\bibitem [{\citenamefont {Micheli}, \citenamefont {Brennen},\ and\
  \citenamefont {Zoller}(2006)}]{Micheli2006}%
  \BibitemOpen
  \bibfield  {author} {\bibinfo {author} {\bibfnamefont {A.}~\bibnamefont
  {Micheli}}, \bibinfo {author} {\bibfnamefont {G.~K.}\ \bibnamefont
  {Brennen}},\ and\ \bibinfo {author} {\bibfnamefont {P.}~\bibnamefont
  {Zoller}},\ }\bibfield  {title} {\enquote {\bibinfo {title} {{A toolbox for
  lattice-spin models with polar molecules}},}\ }\href
  {https://doi.org/10.1038/nphys287} {\bibfield  {journal} {\bibinfo  {journal}
  {Nature Physics}\ }\textbf {\bibinfo {volume} {2}},\ \bibinfo {pages} {341}
  (\bibinfo {year} {2006})}\BibitemShut {NoStop}%
\bibitem [{\citenamefont {Lewenstein}(2006)}]{Lewenstein2006}%
  \BibitemOpen
  \bibfield  {author} {\bibinfo {author} {\bibfnamefont {M.}~\bibnamefont
  {Lewenstein}},\ }\bibfield  {title} {\enquote {\bibinfo {title} {{Polar
  molecules in topological order}},}\ }\href {https://doi.org/10.1038/nphys301}
  {\bibfield  {journal} {\bibinfo  {journal} {Nature Physics}\ }\textbf
  {\bibinfo {volume} {2}},\ \bibinfo {pages} {309} (\bibinfo {year}
  {2006})}\BibitemShut {NoStop}%
\bibitem [{\citenamefont {Guttridge}\ \emph {et~al.}(2018)\citenamefont
  {Guttridge}, \citenamefont {Frye}, \citenamefont {Yang}, \citenamefont
  {Hutson},\ and\ \citenamefont {Cornish}}]{Guttridge2018}%
  \BibitemOpen
  \bibfield  {author} {\bibinfo {author} {\bibfnamefont {A.}~\bibnamefont
  {Guttridge}}, \bibinfo {author} {\bibfnamefont {M.~D.}\ \bibnamefont {Frye}},
  \bibinfo {author} {\bibfnamefont {B.~C.}\ \bibnamefont {Yang}}, \bibinfo
  {author} {\bibfnamefont {J.~M.}\ \bibnamefont {Hutson}},\ and\ \bibinfo
  {author} {\bibfnamefont {S.~L.}\ \bibnamefont {Cornish}},\ }\bibfield
  {title} {\enquote {\bibinfo {title} {{Two-photon photoassociation
  spectroscopy of {CsYb}: Ground-state interaction potential and interspecies
  scattering lengths}},}\ }\href {https://doi.org/10.1103/PHYSREVA.98.022707}
  {\bibfield  {journal} {\bibinfo  {journal} {Physical Review A}\ }\textbf
  {\bibinfo {volume} {98}},\ \bibinfo {pages} {022707} (\bibinfo {year}
  {2018})}\BibitemShut {NoStop}%
\bibitem [{\citenamefont {Green}\ \emph {et~al.}(2019)\citenamefont {Green},
  \citenamefont {{See Toh}}, \citenamefont {Roy}, \citenamefont {Li},
  \citenamefont {Kotochigova},\ and\ \citenamefont {Gupta}}]{Green2019}%
  \BibitemOpen
  \bibfield  {author} {\bibinfo {author} {\bibfnamefont {A.}~\bibnamefont
  {Green}}, \bibinfo {author} {\bibfnamefont {J.~H.}\ \bibnamefont {{See
  Toh}}}, \bibinfo {author} {\bibfnamefont {R.}~\bibnamefont {Roy}}, \bibinfo
  {author} {\bibfnamefont {M.}~\bibnamefont {Li}}, \bibinfo {author}
  {\bibfnamefont {S.}~\bibnamefont {Kotochigova}},\ and\ \bibinfo {author}
  {\bibfnamefont {S.}~\bibnamefont {Gupta}},\ }\bibfield  {title} {\enquote
  {\bibinfo {title} {{Two-photon photoassociation spectroscopy of the
  $^2$$\Sigma$$^+$ YbLi molecular ground state}},}\ }\href
  {https://doi.org/10.1103/PHYSREVA.99.063416} {\bibfield  {journal} {\bibinfo
  {journal} {Physical Review A}\ }\textbf {\bibinfo {volume} {99}},\ \bibinfo
  {pages} {063416} (\bibinfo {year} {2019})}\BibitemShut {NoStop}%
\bibitem [{\citenamefont {Green}\ \emph {et~al.}(2020)\citenamefont {Green},
  \citenamefont {Li}, \citenamefont {{See Toh}}, \citenamefont {Tang},
  \citenamefont {McCormick}, \citenamefont {Li}, \citenamefont {Tiesinga},
  \citenamefont {Kotochigova},\ and\ \citenamefont {Gupta}}]{Green2020}%
  \BibitemOpen
  \bibfield  {author} {\bibinfo {author} {\bibfnamefont {A.}~\bibnamefont
  {Green}}, \bibinfo {author} {\bibfnamefont {H.}~\bibnamefont {Li}}, \bibinfo
  {author} {\bibfnamefont {J.~H.}\ \bibnamefont {{See Toh}}}, \bibinfo {author}
  {\bibfnamefont {X.}~\bibnamefont {Tang}}, \bibinfo {author} {\bibfnamefont
  {K.~C.}\ \bibnamefont {McCormick}}, \bibinfo {author} {\bibfnamefont
  {M.}~\bibnamefont {Li}}, \bibinfo {author} {\bibfnamefont {E.}~\bibnamefont
  {Tiesinga}}, \bibinfo {author} {\bibfnamefont {S.}~\bibnamefont
  {Kotochigova}},\ and\ \bibinfo {author} {\bibfnamefont {S.}~\bibnamefont
  {Gupta}},\ }\bibfield  {title} {\enquote {\bibinfo {title} {{Feshbach
  Resonances in p-Wave Three-Body Recombination within Fermi-Fermi Mixtures of
  Open-Shell $^{6}$Li and Closed-Shell $^{173}$Yb Atoms}},}\ }\href
  {https://doi.org/10.1103/PHYSREVX.10.031037} {\bibfield  {journal} {\bibinfo
  {journal} {Physical Review X}\ }\textbf {\bibinfo {volume} {10}},\ \bibinfo
  {pages} {031037} (\bibinfo {year} {2020})}\BibitemShut {NoStop}%
\bibitem [{\citenamefont {Franzen}\ \emph {et~al.}(2022)\citenamefont
  {Franzen}, \citenamefont {Guttridge}, \citenamefont {Wilson}, \citenamefont
  {Segal}, \citenamefont {Frye}, \citenamefont {Hutson},\ and\ \citenamefont
  {Cornish}}]{Franzen2022}%
  \BibitemOpen
  \bibfield  {author} {\bibinfo {author} {\bibfnamefont {T.}~\bibnamefont
  {Franzen}}, \bibinfo {author} {\bibfnamefont {A.}~\bibnamefont {Guttridge}},
  \bibinfo {author} {\bibfnamefont {K.~E.}\ \bibnamefont {Wilson}}, \bibinfo
  {author} {\bibfnamefont {J.}~\bibnamefont {Segal}}, \bibinfo {author}
  {\bibfnamefont {M.~D.}\ \bibnamefont {Frye}}, \bibinfo {author}
  {\bibfnamefont {J.~M.}\ \bibnamefont {Hutson}},\ and\ \bibinfo {author}
  {\bibfnamefont {S.~L.}\ \bibnamefont {Cornish}},\ }\bibfield  {title}
  {\enquote {\bibinfo {title} {{ Observation of magnetic Feshbach resonances
  between Cs and $^{173}$Yb}},}\ }\href
  {https://doi.org/10.1103/PHYSREVRESEARCH.4.043072} {\bibfield  {journal}
  {\bibinfo  {journal} {Physical Review Research}\ }\textbf {\bibinfo {volume}
  {4}},\ \bibinfo {pages} {043072} (\bibinfo {year} {2022})}\BibitemShut
  {NoStop}%
\bibitem [{\citenamefont {\ifmmode~\dot{Z}\else \.{Z}\fi{}uchowski},
  \citenamefont {Aldegunde},\ and\ \citenamefont
  {Hutson}(2010)}]{Zuchowski2010urs}%
  \BibitemOpen
  \bibfield  {author} {\bibinfo {author} {\bibfnamefont {P.~S.}\ \bibnamefont
  {\ifmmode~\dot{Z}\else \.{Z}\fi{}uchowski}}, \bibinfo {author} {\bibfnamefont
  {J.}~\bibnamefont {Aldegunde}},\ and\ \bibinfo {author} {\bibfnamefont
  {J.~M.}\ \bibnamefont {Hutson}},\ }\bibfield  {title} {\enquote {\bibinfo
  {title} {Ultracold {RbSr} molecules can be formed by magnetoassociation},}\
  }\href {https://doi.org/10.1103/PhysRevLett.105.153201} {\bibfield  {journal}
  {\bibinfo  {journal} {Phys. Rev. Lett.}\ }\textbf {\bibinfo {volume} {105}},\
  \bibinfo {pages} {153201} (\bibinfo {year} {2010})}\BibitemShut {NoStop}%
\bibitem [{\citenamefont {Brue}\ and\ \citenamefont {Hutson}(2012)}]{Brue2012}%
  \BibitemOpen
  \bibfield  {author} {\bibinfo {author} {\bibfnamefont {D.~A.}\ \bibnamefont
  {Brue}}\ and\ \bibinfo {author} {\bibfnamefont {J.~M.}\ \bibnamefont
  {Hutson}},\ }\bibfield  {title} {\enquote {\bibinfo {title} {{Magnetically
  Tunable Feshbach Resonances in Ultracold Li-Yb Mixtures}},}\ }\href
  {https://doi.org/10.1103/PhysRevLett.108.043201} {\bibfield  {journal}
  {\bibinfo  {journal} {Physical Review Letters}\ }\textbf {\bibinfo {volume}
  {108}},\ \bibinfo {pages} {043201} (\bibinfo {year} {2012})}\BibitemShut
  {NoStop}%
\bibitem [{\citenamefont {Brue}\ and\ \citenamefont {Hutson}(2013)}]{Brue2013}%
  \BibitemOpen
  \bibfield  {author} {\bibinfo {author} {\bibfnamefont {D.~A.}\ \bibnamefont
  {Brue}}\ and\ \bibinfo {author} {\bibfnamefont {J.~M.}\ \bibnamefont
  {Hutson}},\ }\bibfield  {title} {\enquote {\bibinfo {title} {{Prospects of
  forming ultracold molecules in $^{2}$$\Sigma$ states by magnetoassociation of
  alkali-metal atoms with Yb}},}\ }\href
  {https://doi.org/10.1103/PhysRevA.87.052709} {\bibfield  {journal} {\bibinfo
  {journal} {Physical Review A}\ }\textbf {\bibinfo {volume} {87}},\ \bibinfo
  {pages} {052709} (\bibinfo {year} {2013})}\BibitemShut {NoStop}%
\bibitem [{\citenamefont {Yang}\ \emph {et~al.}(2019)\citenamefont {Yang},
  \citenamefont {Xie}, \citenamefont {Ji}, \citenamefont {Wang}, \citenamefont
  {Zhang}, \citenamefont {Chen},\ and\ \citenamefont {Jiang}}]{Yang2019uln}%
  \BibitemOpen
  \bibfield  {author} {\bibinfo {author} {\bibfnamefont {Y.-M.}\ \bibnamefont
  {Yang}}, \bibinfo {author} {\bibfnamefont {H.-T.}\ \bibnamefont {Xie}},
  \bibinfo {author} {\bibfnamefont {W.-C.}\ \bibnamefont {Ji}}, \bibinfo
  {author} {\bibfnamefont {Y.-F.}\ \bibnamefont {Wang}}, \bibinfo {author}
  {\bibfnamefont {W.-Y.}\ \bibnamefont {Zhang}}, \bibinfo {author}
  {\bibfnamefont {S.}~\bibnamefont {Chen}},\ and\ \bibinfo {author}
  {\bibfnamefont {X.}~\bibnamefont {Jiang}},\ }\bibfield  {title} {\enquote
  {\bibinfo {title} {Ultra-low noise and high bandwidth bipolar current driver
  for precise magnetic field control},}\ }\href
  {https://doi.org/10.1063/1.5046484} {\bibfield  {journal} {\bibinfo
  {journal} {Review of Scientific Instruments}\ }\textbf {\bibinfo {volume}
  {90}},\ \bibinfo {pages} {014701} (\bibinfo {year} {2019})}\BibitemShut
  {NoStop}%
\bibitem [{\citenamefont {Thomas}\ and\ \citenamefont
  {Kjærgaard}(2020)}]{Thomas2020dig}%
  \BibitemOpen
  \bibfield  {author} {\bibinfo {author} {\bibfnamefont {R.}~\bibnamefont
  {Thomas}}\ and\ \bibinfo {author} {\bibfnamefont {N.}~\bibnamefont
  {Kjærgaard}},\ }\bibfield  {title} {\enquote {\bibinfo {title} {A digital
  feedback controller for stabilizing large electric currents to the ppm level
  for {Feshbach} resonance studies},}\ }\href
  {https://doi.org/10.1063/1.5128935} {\bibfield  {journal} {\bibinfo
  {journal} {Review of Scientific Instruments}\ }\textbf {\bibinfo {volume}
  {91}},\ \bibinfo {pages} {034705} (\bibinfo {year} {2020})}\BibitemShut
  {NoStop}%
\bibitem [{\citenamefont {Xu}\ \emph {et~al.}(2019)\citenamefont {Xu},
  \citenamefont {Wang}, \citenamefont {Jiao}, \citenamefont {Yi}, \citenamefont
  {Sun},\ and\ \citenamefont {Chen}}]{Xu2019ulnmag}%
  \BibitemOpen
  \bibfield  {author} {\bibinfo {author} {\bibfnamefont {X.-T.}\ \bibnamefont
  {Xu}}, \bibinfo {author} {\bibfnamefont {Z.-Y.}\ \bibnamefont {Wang}},
  \bibinfo {author} {\bibfnamefont {R.-H.}\ \bibnamefont {Jiao}}, \bibinfo
  {author} {\bibfnamefont {C.-R.}\ \bibnamefont {Yi}}, \bibinfo {author}
  {\bibfnamefont {W.}~\bibnamefont {Sun}},\ and\ \bibinfo {author}
  {\bibfnamefont {S.}~\bibnamefont {Chen}},\ }\bibfield  {title} {\enquote
  {\bibinfo {title} {Ultra-low noise magnetic field for quantum gases},}\
  }\href {https://doi.org/10.1063/1.5087957} {\bibfield  {journal} {\bibinfo
  {journal} {Review of Scientific Instruments}\ }\textbf {\bibinfo {volume}
  {90}},\ \bibinfo {pages} {054708} (\bibinfo {year} {2019})}\BibitemShut
  {NoStop}%
\bibitem [{\citenamefont {Merkel}\ \emph {et~al.}(2019)\citenamefont {Merkel},
  \citenamefont {Thirumalai}, \citenamefont {Tarlton}, \citenamefont
  {Schäfer}, \citenamefont {Ballance}, \citenamefont {Harty},\ and\
  \citenamefont {Lucas}}]{Merkel2019magstab}%
  \BibitemOpen
  \bibfield  {author} {\bibinfo {author} {\bibfnamefont {B.}~\bibnamefont
  {Merkel}}, \bibinfo {author} {\bibfnamefont {K.}~\bibnamefont {Thirumalai}},
  \bibinfo {author} {\bibfnamefont {J.~E.}\ \bibnamefont {Tarlton}}, \bibinfo
  {author} {\bibfnamefont {V.~M.}\ \bibnamefont {Schäfer}}, \bibinfo {author}
  {\bibfnamefont {C.~J.}\ \bibnamefont {Ballance}}, \bibinfo {author}
  {\bibfnamefont {T.~P.}\ \bibnamefont {Harty}},\ and\ \bibinfo {author}
  {\bibfnamefont {D.~M.}\ \bibnamefont {Lucas}},\ }\bibfield  {title} {\enquote
  {\bibinfo {title} {Magnetic field stabilization system for atomic physics
  experiments},}\ }\href {https://doi.org/10.1063/1.5080093} {\bibfield
  {journal} {\bibinfo  {journal} {Review of Scientific Instruments}\ }\textbf
  {\bibinfo {volume} {90}},\ \bibinfo {pages} {044702} (\bibinfo {year}
  {2019})}\BibitemShut {NoStop}%
\bibitem [{\citenamefont {Duan}\ \emph {et~al.}(2022)\citenamefont {Duan},
  \citenamefont {Wu}, \citenamefont {Lin},\ and\ \citenamefont
  {Yang}}]{Duan2022}%
  \BibitemOpen
  \bibfield  {author} {\bibinfo {author} {\bibfnamefont {Z.-X.}\ \bibnamefont
  {Duan}}, \bibinfo {author} {\bibfnamefont {W.-T.}\ \bibnamefont {Wu}},
  \bibinfo {author} {\bibfnamefont {Y.-T.}\ \bibnamefont {Lin}},\ and\ \bibinfo
  {author} {\bibfnamefont {S.-J.}\ \bibnamefont {Yang}},\ }\bibfield  {title}
  {\enquote {\bibinfo {title} {{Simple and active magnetic-field stabilization
  for cold atom experiments}},}\ }\href {https://doi.org/10.1063/5.0119778}
  {\bibfield  {journal} {\bibinfo  {journal} {Review of Scientific
  Instruments}\ }\textbf {\bibinfo {volume} {93}},\ \bibinfo {pages} {123201}
  (\bibinfo {year} {2022})}\BibitemShut {NoStop}%
\bibitem [{\citenamefont {O'Dwyer}\ \emph {et~al.}(2020)\citenamefont
  {O'Dwyer}, \citenamefont {Ingleby}, \citenamefont {Chalmers}, \citenamefont
  {Griffin},\ and\ \citenamefont {Riis}}]{ODwyer2020}%
  \BibitemOpen
  \bibfield  {author} {\bibinfo {author} {\bibfnamefont {C.}~\bibnamefont
  {O'Dwyer}}, \bibinfo {author} {\bibfnamefont {S.~J.}\ \bibnamefont
  {Ingleby}}, \bibinfo {author} {\bibfnamefont {I.~C.}\ \bibnamefont
  {Chalmers}}, \bibinfo {author} {\bibfnamefont {P.~F.}\ \bibnamefont
  {Griffin}},\ and\ \bibinfo {author} {\bibfnamefont {E.}~\bibnamefont
  {Riis}},\ }\bibfield  {title} {\enquote {\bibinfo {title} {{A feed-forward
  measurement scheme for periodic noise suppression in atomic magnetometry}},}\
  }\href {https://doi.org/10.1063/5.0002964} {\bibfield  {journal} {\bibinfo
  {journal} {Review of Scientific Instruments}\ }\textbf {\bibinfo {volume}
  {91}},\ \bibinfo {pages} {045103} (\bibinfo {year} {2020})}\BibitemShut
  {NoStop}%
\bibitem [{\citenamefont {Pyragius}\ and\ \citenamefont
  {Jensen}(2021)}]{Pyragius2021}%
  \BibitemOpen
  \bibfield  {author} {\bibinfo {author} {\bibfnamefont {T.}~\bibnamefont
  {Pyragius}}\ and\ \bibinfo {author} {\bibfnamefont {K.}~\bibnamefont
  {Jensen}},\ }\bibfield  {title} {\enquote {\bibinfo {title} {{A high
  performance active noise control system for magnetic fields}},}\ }\href
  {https://doi.org/10.1063/5.0062650} {\bibfield  {journal} {\bibinfo
  {journal} {Review of Scientific Instruments}\ }\textbf {\bibinfo {volume}
  {92}},\ \bibinfo {pages} {124702} (\bibinfo {year} {2021})}\BibitemShut
  {NoStop}%
\bibitem [{\citenamefont {Wei}\ \emph {et~al.}(2022)\citenamefont {Wei},
  \citenamefont {Hao}, \citenamefont {Ma}, \citenamefont {Zhang}, \citenamefont
  {Pang}, \citenamefont {Wu}, \citenamefont {Deng}, \citenamefont {Zhang},\
  and\ \citenamefont {Lu}}]{Wei2022}%
  \BibitemOpen
  \bibfield  {author} {\bibinfo {author} {\bibfnamefont {W.}~\bibnamefont
  {Wei}}, \bibinfo {author} {\bibfnamefont {P.}~\bibnamefont {Hao}}, \bibinfo
  {author} {\bibfnamefont {Z.}~\bibnamefont {Ma}}, \bibinfo {author}
  {\bibfnamefont {H.}~\bibnamefont {Zhang}}, \bibinfo {author} {\bibfnamefont
  {L.}~\bibnamefont {Pang}}, \bibinfo {author} {\bibfnamefont {F.}~\bibnamefont
  {Wu}}, \bibinfo {author} {\bibfnamefont {K.}~\bibnamefont {Deng}}, \bibinfo
  {author} {\bibfnamefont {J.}~\bibnamefont {Zhang}},\ and\ \bibinfo {author}
  {\bibfnamefont {Z.}~\bibnamefont {Lu}},\ }\bibfield  {title} {\enquote
  {\bibinfo {title} {{Measurement and suppression of magnetic field noise of
  trapped ion qubit}},}\ }\href {https://doi.org/10.1088/1361-6455/AC5E7D}
  {\bibfield  {journal} {\bibinfo  {journal} {Journal of Physics B: Atomic,
  Molecular and Optical Physics}\ }\textbf {\bibinfo {volume} {55}},\ \bibinfo
  {pages} {075001} (\bibinfo {year} {2022})}\BibitemShut {NoStop}%
\bibitem [{\citenamefont {Schreck}\ and\ \citenamefont
  {Meyrath}(2011)}]{Schreck2011control}%
  \BibitemOpen
  \bibfield  {author} {\bibinfo {author} {\bibfnamefont {F.}~\bibnamefont
  {Schreck}}\ and\ \bibinfo {author} {\bibfnamefont {T.}~\bibnamefont
  {Meyrath}},\ }\href {http://www.strontiumbec.com/} {\enquote {\bibinfo
  {title} {Ultracold atom experiment control system {CONTROL}},}\ } (\bibinfo
  {year} {2011})\BibitemShut {NoStop}%
\bibitem [{\citenamefont {Wille}(2009)}]{Wille2009poo}%
  \BibitemOpen
  \bibfield  {author} {\bibinfo {author} {\bibfnamefont {E.}~\bibnamefont
  {Wille}},\ }\emph {\bibinfo {title} {Preparation of an Optically Trapped
  Fermi-Fermi Mixture of $^6$Li and $^{40}$K Atoms and Characterization of the
  Interspecies Interactions by {Feshbach} Spectroscopy}},\ \href
  {http://www.ultracold.at/theses/2009-wille.pdf} {Ph.D. thesis},\ \bibinfo
  {school} {Innsbruck University} (\bibinfo {year} {2009})\BibitemShut
  {NoStop}%
\bibitem [{\citenamefont {Horowitz}\ and\ \citenamefont
  {Hill}(2015)}]{horowitz2015art}%
  \BibitemOpen
  \bibfield  {author} {\bibinfo {author} {\bibfnamefont {P.}~\bibnamefont
  {Horowitz}}\ and\ \bibinfo {author} {\bibfnamefont {W.}~\bibnamefont
  {Hill}},\ }\href@noop {} {\emph {\bibinfo {title} {The Art of
  Electronics}}},\ \bibinfo {edition} {3rd}\ ed.\ (\bibinfo  {publisher}
  {Cambridge University Press},\ \bibinfo {address} {USA},\ \bibinfo {year}
  {2015})\BibitemShut {NoStop}%
\bibitem [{\citenamefont {Miller}(1920)}]{Miller1920dot}%
  \BibitemOpen
  \bibfield  {author} {\bibinfo {author} {\bibfnamefont {J.~M.}\ \bibnamefont
  {Miller}},\ }\bibfield  {title} {\enquote {\bibinfo {title} {Dependence of
  the input impedance of a three-electrode vacuum tube upon the load in the
  plate circuit},}\ }\href {https://doi.org/10.6028/nbsscipaper.024} {\bibfield
   {journal} {\bibinfo  {journal} {Scientific Papers of the Bureau of
  Standards}\ }\textbf {\bibinfo {volume} {15}},\ \bibinfo {pages} {367}
  (\bibinfo {year} {1920})}\BibitemShut {NoStop}%
\bibitem [{\citenamefont {Mies}, \citenamefont {Tiesinga},\ and\ \citenamefont
  {Julienne}(2000)}]{Mies2000}%
  \BibitemOpen
  \bibfield  {author} {\bibinfo {author} {\bibfnamefont {F.~H.}\ \bibnamefont
  {Mies}}, \bibinfo {author} {\bibfnamefont {E.}~\bibnamefont {Tiesinga}},\
  and\ \bibinfo {author} {\bibfnamefont {P.~S.}\ \bibnamefont {Julienne}},\
  }\bibfield  {title} {\enquote {\bibinfo {title} {{Manipulation of Feshbach
  resonances in ultracold atomic collisions using time-dependent magnetic
  fields}},}\ }\href {https://doi.org/10.1103/PhysRevA.61.022721} {\bibfield
  {journal} {\bibinfo  {journal} {Physical Review A}\ }\textbf {\bibinfo
  {volume} {61}},\ \bibinfo {pages} {022721} (\bibinfo {year}
  {2000})}\BibitemShut {NoStop}%
\bibitem [{\citenamefont {G{\'{o}}ral}\ \emph {et~al.}(2004)\citenamefont
  {G{\'{o}}ral}, \citenamefont {K{\"{o}}hler}, \citenamefont {Gardiner},
  \citenamefont {Tiesinga},\ and\ \citenamefont {Julienne}}]{Goral2004}%
  \BibitemOpen
  \bibfield  {author} {\bibinfo {author} {\bibfnamefont {K.}~\bibnamefont
  {G{\'{o}}ral}}, \bibinfo {author} {\bibfnamefont {T.}~\bibnamefont
  {K{\"{o}}hler}}, \bibinfo {author} {\bibfnamefont {S.~A.}\ \bibnamefont
  {Gardiner}}, \bibinfo {author} {\bibfnamefont {E.}~\bibnamefont {Tiesinga}},\
  and\ \bibinfo {author} {\bibfnamefont {P.~S.}\ \bibnamefont {Julienne}},\
  }\bibfield  {title} {\enquote {\bibinfo {title} {{Adiabatic association of
  ultracold molecules via magnetic-field tunable interactions}},}\ }\href
  {https://doi.org/10.1088/0953-4075/37/17/006} {\bibfield  {journal} {\bibinfo
   {journal} {Journal of Physics B: Atomic, Molecular and Optical Physics}\
  }\textbf {\bibinfo {volume} {37}},\ \bibinfo {pages} {3457} (\bibinfo {year}
  {2004})}\BibitemShut {NoStop}%
\bibitem [{\citenamefont {Pasquiou}\ \emph {et~al.}(2013)\citenamefont
  {Pasquiou}, \citenamefont {Bayerle}, \citenamefont {Tzanova}, \citenamefont
  {Stellmer}, \citenamefont {Szczepkowski}, \citenamefont {Parigger},
  \citenamefont {Grimm},\ and\ \citenamefont {Schreck}}]{Pasquiou2013qdm}%
  \BibitemOpen
  \bibfield  {author} {\bibinfo {author} {\bibfnamefont {B.}~\bibnamefont
  {Pasquiou}}, \bibinfo {author} {\bibfnamefont {A.}~\bibnamefont {Bayerle}},
  \bibinfo {author} {\bibfnamefont {S.~M.}\ \bibnamefont {Tzanova}}, \bibinfo
  {author} {\bibfnamefont {S.}~\bibnamefont {Stellmer}}, \bibinfo {author}
  {\bibfnamefont {J.}~\bibnamefont {Szczepkowski}}, \bibinfo {author}
  {\bibfnamefont {M.}~\bibnamefont {Parigger}}, \bibinfo {author}
  {\bibfnamefont {R.}~\bibnamefont {Grimm}},\ and\ \bibinfo {author}
  {\bibfnamefont {F.}~\bibnamefont {Schreck}},\ }\bibfield  {title} {\enquote
  {\bibinfo {title} {Quantum degenerate mixtures of strontium and rubidium
  atoms},}\ }\href {https://doi.org/10.1103/PhysRevA.88.023601} {\bibfield
  {journal} {\bibinfo  {journal} {Phys. Rev. A}\ }\textbf {\bibinfo {volume}
  {88}},\ \bibinfo {pages} {023601} (\bibinfo {year} {2013})}\BibitemShut
  {NoStop}%
\bibitem [{\citenamefont {Harber}\ \emph {et~al.}(2002)\citenamefont {Harber},
  \citenamefont {Lewandowski}, \citenamefont {McGuirk},\ and\ \citenamefont
  {Cornell}}]{Harber2002}%
  \BibitemOpen
  \bibfield  {author} {\bibinfo {author} {\bibfnamefont {D.~M.}\ \bibnamefont
  {Harber}}, \bibinfo {author} {\bibfnamefont {H.~J.}\ \bibnamefont
  {Lewandowski}}, \bibinfo {author} {\bibfnamefont {J.~M.}\ \bibnamefont
  {McGuirk}},\ and\ \bibinfo {author} {\bibfnamefont {E.~A.}\ \bibnamefont
  {Cornell}},\ }\bibfield  {title} {\enquote {\bibinfo {title} {Effect of cold
  collisions on spin coherence and resonance shifts in a magnetically trapped
  ultracold gas},}\ }\href {https://doi.org/10.1103/PhysRevA.66.053616}
  {\bibfield  {journal} {\bibinfo  {journal} {Phys. Rev. A}\ }\textbf {\bibinfo
  {volume} {66}},\ \bibinfo {pages} {053616} (\bibinfo {year}
  {2002})}\BibitemShut {NoStop}%
\bibitem [{\citenamefont {Borkowski}\ \emph {et~al.}(2023)\citenamefont
  {Borkowski}, \citenamefont {Reichs{\"{o}}llner}, \citenamefont {Thekkeppatt},
  \citenamefont {Barb{\'{e}}}, \citenamefont {{Van Roon}}, \citenamefont {van
  Druten},\ and\ \citenamefont {Schreck}}]{magnetic_field_schematics}%
  \BibitemOpen
  \bibfield  {author} {\bibinfo {author} {\bibfnamefont {M.}~\bibnamefont
  {Borkowski}}, \bibinfo {author} {\bibfnamefont {L.}~\bibnamefont
  {Reichs{\"{o}}llner}}, \bibinfo {author} {\bibfnamefont {P.}~\bibnamefont
  {Thekkeppatt}}, \bibinfo {author} {\bibfnamefont {V.}~\bibnamefont
  {Barb{\'{e}}}}, \bibinfo {author} {\bibfnamefont {T.}~\bibnamefont {{Van
  Roon}}}, \bibinfo {author} {\bibfnamefont {K.}~\bibnamefont {van Druten}},\
  and\ \bibinfo {author} {\bibfnamefont {F.}~\bibnamefont {Schreck}},\ }\href
  {https://github.com/StrontiumGroup/Magnetic-field-contraption} {\enquote
  {\bibinfo {title} {{Magnetic field stabilization system schematics}},}\ }
  (\bibinfo {year} {2023})\BibitemShut {NoStop}%
\end{thebibliography}%

\end{document}